\documentstyle[12pt]{article}
\newcommand{\nc}{\newcommand}
\nc{\renc}{\renewcommand}

\nc{\half}{{\textstyle{1\over2}}}
\nc{\etal}{\mbox{\it et al. }}
\nc{\ie}{{\it i.e.}}
\nc{\eg}{{\it e.g.}}

\renc{\thefootnote}{\arabic{footnote}}
\nc{\capt}[1]{{\bf Figure.} {\small\sl #1}}

\nc{\eqs}[2]{\mbox{Eqs.~(\ref{#1},\,\ref{#2})}}
\nc{\eq}[1]{\mbox{Eq.~(\ref{#1})}}

\nc{\figs}[2]{\mbox{Figs.~(\ref{#1},\,\ref{#2})}}
\nc{\fig}[1]{\mbox{Fig~.(\ref{#1})}}

\nc{\tag}[1]{\label{#1} \marginpar{{\footnotesize #1}}}
\nc{\mtag}[1]{\label{#1} \mbox{\marginpar{{\footnotesize #1}}}}
\renc{\baselinestretch}{1.5}
\newlength{\overeqskip}
\newlength{\undereqskip}
\nc{\be}[1]{\begin{equation} \mbox{$\label{#1}$}}
\nc{\bea}[1]{\begin{eqnarray} \mbox{$\label{#1}$}}
\nc{\Section}[2]{\section{#2}\label{#1}}
\nc{\Bibitem}[1]{\bibitem{#1}}
\nc{\Label}[1]{\label{#1}}
\nc{\eea}{\vspace{\undereqskip}\end{eqnarray}}
\nc{\ee}{\vspace{\undereqskip}\end{equation}}
\nc{\bdm}{\begin{displaymath}}
\nc{\edm}{\end{displaymath}}
\nc{\dpsty}{\displaystyle}
\nc{\bc}{\begin{center}}
\nc{\ec}{\end{center}}
\nc{\ba}{\begin{array}}
\nc{\ea}{\end{array}}
\nc{\bab}{\begin{abstract}}
\nc{\eab}{\end{abstract}}
\nc{\btab}{\begin{tabular}}
\nc{\etab}{\end{tabular}}
\nc{\bit}{\begin{itemize}}
\nc{\eit}{\end{itemize}}
\nc{\ben}{\begin{enumerate}}
\nc{\een}{\end{enumerate}}
\nc{\bfig}{\begin{figure}}
\nc{\efig}{\end{figure}}
\nc{\arreq}{&\!=\!&}
\nc{\arrmi}{&\!-\!&}
\nc{\arrpl}{&\!+\!&}
\nc{\arrap}{&\!\!\!\approx\!\!\!&}
\nc{\non}{\nonumber\\*}
\nc{\align}{\!\!\!\!\!\!\!\!&&}

\def\lsim{\; \raise0.3ex\hbox{$<$\kern-0.75em
      \raise-1.1ex\hbox{$\sim$}}\; }
\def\gsim{\; \raise0.3ex\hbox{$>$\kern-0.75em
      \raise-1.1ex\hbox{$\sim$}}\; }
\nc{\DOT}{\hspace{-0.08in}{\bf .}\hspace{0.1in}}
\nc{\Laada}{\hbox {$\sqcap$ \kern -1em $\sqcup$}}
\nc\loota{{\scriptstyle\sqcap\kern-0.55em\hbox{$\scriptstyle\sqcup$}}}
\nc\Loota{{\sqcap\kern-0.65em\hbox{$\sqcup$}}}
\nc\laada{\Loota}
\nc{\qed}{\hskip 3em \hbox{\BOX} \vskip 2ex}

\nc{\real}{{\rm I \! R}}
\nc{\Z}{{\sf Z \!\!\! Z}}
\nc{\complex}{{\rm C\!\!\! {\sf I}\,\,}}
\def\bigid{\leavevmode\hbox{\small1\kern-3.8pt\normalsize1}}
\def\id{\leavevmode\hbox{\small1\kern-3.3pt\normalsize1}}
\nc{\slask}{\!\!\!/}
\nc{\bis}{{\prime\prime}}
\nc{\pa}{\partial}
\nc{\na}{\nabla}
\nc{\ra}{\rangle}
\nc{\la}{\langle}
\nc{\goto}{\rightarrow}
\nc{\swap}{\leftrightarrow}

\nc{\EE}[1]{ \mbox{$\cdot10^{#1}$} }
\nc{\abs}[1]{\left|#1\right|}
\nc{\at}[2]{\left.#1\right|_{#2}}
\nc{\norm}[1]{\|#1\|}
\nc{\abscut}[2]{\Abs{#1}_{\scriptscriptstyle#2}}
\nc{\vek}[1]{{\rm\bf #1}}
\nc{\integral}[2]{\int\limits_{#1}^{#2}}
\nc{\inv}[1]{\frac{1}{#1}}
\nc{\dd}[2]{{{\partial #1}\over{\partial #2}}}
\nc{\ddd}[2]{{{{\partial}^2 #1}\over{\partial {#2}^2}}}
\nc{\dddd}[3]{{{{\partial}^2 #1}\over
	{\partial #2 \partial #3}}}
\nc{\dder}[2]{{{d #1}\over{d #2}}}
\nc{\ddder}[2]{{{d^2 #1}\over{d {#2}^2}}}
\nc{\dddder}[3]{{d^2 #1}\over
	{d #2 d #3}}
\nc{\dx}[1]{d\,^{#1}x}
\nc{\dy}[1]{d\,^{#1}y}
\nc{\dz}[1]{d\,^{#1}z}
\nc{\dl}[1]{\frac{d\,^{#1}l}{(2\pi)^{#1}}}
\nc{\dk}[1]{\frac{d\,^{#1}k}{(2\pi)^{#1}}}
\nc{\dq}[1]{\frac{d\,^{#1}q}{(2\pi)^{#1}}}

\nc{\cc}{\mbox{$c.c.$ }}
\nc{\hc}{\mbox{$h.c.$ }}
\nc{\cf}{cf.\ }
\nc{\erfc}{{\rm erfc}}
\nc{\Tr}{{\rm Tr\,}}
\nc{\tr}{{\rm tr\,}}
\nc{\pol}{{\rm pol}}
\nc{\sign}{{\rm sign}}
\nc{\bfT}{{\bf T }}

\def\GeV{{\rm\ GeV}}

\nc{\cA}{{\cal A}}
\nc{\cB}{{\cal B}}
\nc{\cD}{{\cal D}}
\nc{\cE}{{\cal E}}
\nc{\cG}{{\cal G}}
\nc{\cH}{{\cal H}}
\nc{\cL}{{\cal L}}
\nc{\cO}{{\cal O}}
\nc{\cT}{{\cal T}}
\nc{\cN}{{\cal N}}
\nc{\rvac}[1]{|{\cal O}#1\rangle}
\nc{\lvac}[1]{\langle{\cal O}#1|}
\nc{\rvacb}[1]{|{\cal O}_\beta #1\rangle}
\nc{\lvacb}[1]{\langle{\cal O}_\beta #1 |}
\nc{\bb}{\bar{\beta}}
\nc{\bt}{\tilde{\beta}}
\nc{\ctH}{\tilde{\cal H}}
\nc{\chH}{\hat{\cal H}}
\nc{\1}{\aa}
\nc{\2}{\"{a}}
\nc{\3}{\"{o}}
\nc{\4}{\AA}
\nc{\5}{\"{A}}
\nc{\6}{\"{O}}
\nc{\al}{\alpha}
\nc{\g}{\gamma}
\nc{\Del}{\Delta}
\nc{\e}{\epsilon}
\nc{\eps}{\epsilon}
\nc{\lam}{\lambda}
\nc{\om}{\omega}
\nc{\Om}{\Omega}
\nc{\ve}{\varepsilon}
\nc{\mn}{{\mu\nu}}
\nc{\k}{\kappa}
\nc{\vp}{\varphi}

\nc{\advp}[3]{{\it  Adv.\ in\ Phys.\ }{{\bf #1} {(#2)} {#3}}}
\nc{\annp}[3]{{\it  Ann.\ Phys.\ (N.Y.)\ }{{\bf #1} {(#2)} {#3}}}
\nc{\apl}[3]{{\it  Appl. Phys. Lett. }{{\bf #1} {(#2)} {#3}}}
\nc{\apj}[3]{{\it  Ap.\ J.\ }{{\bf #1} {(#2)} {#3}}}
\nc{\apjl}[3]{{\it  Ap.\ J.\ Lett.\ }{{\bf #1} {(#2)} {#3}}}
\nc{\app}[3]{{\it Astropart.\ Phys.\ }{{\bf #1} {(#2)} {#3}}}
\nc{\cmp}[3]{{\it  Comm.\ Math.\ Phys.\ }{{ \bf #1} {(#2)} {#3}}}
\nc{\cqg}[3]{{\it  Class.\ Quant.\ Grav.\ }{{\bf #1} {(#2)} {#3}}}
\nc{\epl}[3]{{\it  Europhys.\ Lett.\ }{{\bf #1} {(#2)} {#3}}}
\nc{\ijmp}[3]{{\it Int.\ J.\ Mod.\ Phys.\ }{{\bf #1} {(#2)} {#3}}}
\nc{\ijtp}[3]{{\it Int.\ J.\ Theor.\ Phys.\ }{{\bf #1} {(#2)} {#3}}}
\nc{\jmp}[3]{{\it  J.\ Math.\ Phys.\ }{{ \bf #1} {(#2)} {#3}}}
\nc{\jpa}[3]{{\it  J.\ Phys.\ A\ }{{\bf #1} {(#2)} {#3}}}
\nc{\jpc}[3]{{\it  J.\ Phys.\ C\ }{{\bf #1} {(#2)} {#3}}}
\nc{\jap}[3]{{\it J.\ Appl.\ Phys.\ }{{\bf #1} {(#2)} {#3}}}
\nc{\jpsj}[3]{{\it J.\ Phys.\ Soc.\ Japan\ }{{\bf #1} {(#2)} {#3}}}
\nc{\lmp}[3]{{\it Lett.\ Math.\ Phys.\ }{{\bf #1} {(#2)} {#3}}}
\nc{\mpl}[3]{{\it  Mod.\ Phys.\ Lett.\ }{{\bf #1} {(#2)} {#3}}}
\nc{\ncim}[3]{{\it  Nuov.\ Cim.\ }{{\bf #1} {(#2)} {#3}}}
\nc{\np}[3]{{\it  Nucl.\ Phys.\ }{{\bf #1} {(#2)} {#3}}}
\nc{\npps}[3]{{\it  Nucl.\ Phys.\ Proc.\ Suppl.\ }{{\bf #1} {(#2)} {#3}}}
\nc{\pr}[3]{{\it Phys.\ Rev.\ }{{\bf #1} {(#2)} {#3}}}
\nc{\pra}[3]{{\it  Phys.\ Rev.\ A\ }{{\bf #1} {(#2)} {#3}}}
\nc{\prb}[3]{{\it  Phys.\ Rev.\ B\ }{{{\bf #1} {(#2)} {#3}}}}
\nc{\prc}[3]{{\it  Phys.\ Rev.\ C\ }{{\bf #1} {(#2)} {#3}}}
\nc{\prd}[3]{{\it  Phys.\ Rev.\ D\ }{{\bf #1} {(#2)} {#3}}}
\nc{\prl}[3]{{\it Phys.\ Rev.\ Lett.\ }{{\bf #1} {(#2)} {#3}}}
\nc{\pl}[3]{{\it  Phys.\ Lett.\ }{{\bf #1} {(#2)} {#3}}}
\nc{\prep}[3]{{\it Phys.\ Rep.\ }{{\bf #1} {(#2)} {#3}}}
\nc{\prsl}[3]{{\it Proc.\ R.\ Soc.\ London\ }{{\bf #1} {(#2)} {#3}}}
\nc{\ptp}[3]{{\it  Prog.\ Theor.\ Phys.\ }{{\bf #1} {(#2)} {#3}}}
\nc{\ptps}[3]{{\it  Prog\ Theor.\ Phys.\ suppl.\ }{{\bf #1} {(#2)} {#3}}}
\nc{\physa}[3]{{\it  Physica\ A\ }{{\bf #1} {(#2)} {#3}}}
\nc{\physb}[3]{{\it  Physica\ B\ }{{\bf #1} {(#2)} {#3}}}
\nc{\phys}[3]{{\it Physica\ }{{\bf #1} {(#2)} {#3}}}
\nc{\rmp}[3]{{\it  Rev.\ Mod.\ Phys.\ }{{\bf #1} {(#2)} {#3}}}
\nc{\rpp}[3]{{\it Rep.\ Prog.\ Phys.\ }{{\bf #1} {(#2)} {#3}}}
\nc{\sjnp}[3]{{\it Sov.\ J.\ Nucl.\ Phys.\ }{{\bf #1} {(#2)} {#3}}}
\nc{\spjetp}[3]{{\it Sov.\ Phys.\ JETP\ }{{\bf #1} {(#2)} {#3}}}
\nc{\yf}[3]{{\it Yad.\ Fiz.\ }{{\bf #1} {(#2)} {#3}}}
\nc{\zetp}[3]{{\it Zh.\ Eksp.\ Teor.\ Fiz.\  }{{\bf #1}  {(#2)} {#3}}}
\nc{\zp}[3]{{\it Z.\ Phys.\ }{{\bf #1} {(#2)} {#3}}}
\nc{\ibid}[3]{{\sl ibid.\ }{{\bf #1} {#2} {#3}}}
\nc{\rf}[1]{(\ref{#1})}
\nc{\nn}{\nonumber \\*}
\nc{\bfB}{\bf{B}}
\nc{\bfv}{\bf{v}}
\nc{\bfx}{\bf{x}}
\nc{\bfy}{\bf{y}}
\nc{\vx}{\vec{x}}
\nc{\vy}{\vec{y}}
\nc{\oB}{\overline{B}}
\nc{\oI}{\overline{I}}
\nc{\oR}{\overline{R}}
\nc{\rar}{\rightarrow}
\nc{\ti}{\times}
\nc{\slsh}{\hskip-5pt/}
\nc{\sm}{Standard~Model~}
\nc{\MP}{M_{\rm Pl}}
\nc{\tp}{t_{\rm Pl}}
\nc{\ave}{\bar{E}}

\nc{\eff}{{\rm eff}}
\nc{\kk}{\vek{k}}
\nc{\pp}{{\rm p}}
\nc{\ga}{g_{a\gamma}}
\nc{\vv}{\\}
\nc{\eee}{{\bf E}}
\nc{\bbb}{{\bf B}}
\nc{\qcd}{T_{\rm QCD}}
\nc{\G}{\rm \ G}
\def\vec#1{{\bf #1}}
\def\lae{\;^{<}_{\sim} \;} \def\gae{\; ^{>}_{\sim} \;} 

\def\udd{u^{c}d^{c}d^{c}}

\def\uude{u^{c}u^{c}d^{c}e^{c}}

\begin{document}
{\title{\vskip-2truecm{\hfill {{\small \\
	\hfill HIP-1998-18/ph \\
	}}\vskip 1truecm}
{\bf  B-ball Baryogenesis and the Baryon to Dark Matter Ratio.}}
%\vspace{1.2cm}
{\author{
{\sc  Kari Enqvist$^{1}$}\\
{\sl\small Department of Physics and Helsinki Institute of Physics,}\\ 
{\sl\small P.O. Box 9,
FIN-00014 University of Helsinki,
Finland}\\
{\sc and}\\
{\sc  John McDonald$^{2}$}\\
{\sl\small Department of Physics, P.O. Box 9,
FIN-00014 University of Helsinki,
Finland}
}
\maketitle
\vspace{1cm}
%\newpage
\begin{abstract}
\noindent

We demonstrate that B-ball decay in the MSSM 
can naturally solve the puzzle of why the densities of 
baryons and dark matter in the Universe are similar. 
 This requires that the B-balls
survive thermalization and decay below the 
freeze-out temperature of the neutralino LSP, typically 1-10 GeV.
It is shown that this can happen if the baryon
asymmetry originates from a squark condensate along the 
d=6 $\udd$ D-flat direction of the MSSM scalar potential. 
For this to work the reheating temperature after inflation 
must be no greater than $10^{3-5}\GeV$. 
\end{abstract}
\vfil
\footnoterule
{\small $^1$enqvist@pcu.helsinki.fi};
{\small $^2$mcdonald@rock.helsinki.fi}

\thispagestyle{empty}
\newpage
\setcounter{page}{1}

%%%%%%%%%%%%%%%%%%%%%%%%%%%%%%%%%%%%%%%%%%%%
%%%%%%%%%%%%%%%%%%%%%%%%%%%%%%%%%%
%%%%%%%%%%%%%%%%%%%%%%%%%%%%%%%%%%%%%%%%%%%
%%%%%%%%%%%%%%%%%%%%%%%%%%%%%%%%%%%
\section{Introduction}

		  Observations of the dynamics of galaxies and clusters of galaxies, as well as theoretical
models of galaxy formation from primordial density fluctuations, all point to a Universe mainly
 composed of non-baryonic dark matter \cite{eu}. Given that, in most models of baryogenesis, the 
physics of dark matter is completely unconnected with the physics of baryogenesis, it is  
remarkable that the ratio of matter in baryons is so close to that in dark matter, 0.5 to 10 
$\%$ in a flat Universe \cite{sarkar}. When one considers that the dark matter particles in weakly interacting 
cold dark matter models,
such as supersymmetric models with unbroken R-parity \cite{susydm}, are often 
heavier that the nucleons, this points to an even closer coincidence when expressed in terms 
of number density. Unless one can produce a convincing anthropic principle 
explanation of
 the coincidence of the baryon to dark matter ratio (such as could naturally 
occur in axion models \cite{lax}), 
this strongly suggests that  the 
origin
 of baryon number 
and dark matter are closely connected. 

	   Electroweak baryogenesis in the
 Minimal Supersymmetric Standard Model (MSSM) and its
 extensions \cite{ewb1,ewb2,ewb3} has some possibility of a connection, in the sense that both baryon number and
 dark matter have their origin in weak interaction physics. However, the connection between the physics of
 dark matter freeze-out \cite{susydm} and that of anomalous B+L violation during the electroweak phase transition
 is at best tenuous, allowing a wide range of possible baryon to dark matter ratios. In this paper we
 will discuss an equally plausible model for baryogenesis, requiring no new
 low energy physics beyond the MSSM, which has the great advantage of being able to
 account for the baryon to dark matter ratio naturally, namely B-Ball Baryogenesis (BBB) \cite{qb1}. 
 This is closely
 related to Affleck-Dine (AD) baryogenesis \cite{ad}, in which a 
B violating scalar condensate forms along a D-flat direction of the MSSM scalar potential composed
of squark and possibly slepton fields. The difference in BBB is that the condensate is naturally
 unstable with respect
 to the formation of Q-balls \cite{cole,cole2,k} of baryon number \cite{ks}. (In the following we will use the term B-ball 
to refer to any Q-ball carrying baryon number). 
The condensate therefore breaks up into 
a mixture of B-balls and free squarks carrying baryon number. 

		      The subsequent evolution of the B-balls will depend crucially upon the scalar potential
 associated with the condensate scalar, which in turn depends upon the SUSY breaking mechanism.
One possibility being vigourously pursued \cite{ks,ksdm,sw,dks} is that SUSY breaking occurs at low energy scales, via gauge
 mediated SUSY breaking \cite{gmsb}. As a result, for field values larger than the mass of the messenger fields,
 the scalar potential of the condensate scalar is completely flat, resulting in B-balls whose energy per unit charge
is proportional to $B^{-1/4}$ \cite{dks}. For large enough $B$, the B-ball cannot decay to the lightest B-carrying fermions
(the nucleons) and so is completely stable. Stable B-balls could have a wide range of astrophysical \cite{sw,ksdm,ks}, 
experimental \cite{ksdm,dks} and practical \cite{dks} implications. 

	    A very different scenario, which is the 
subject of the present paper, emerges if SUSY 
breaking occurs via the more conventional supergravity hidden sector breaking
 mechanism \cite{nilles}. In this case the potential
is not flat, but nevertheless radiative corrections to the $\phi^{2}$-type
 condensate potential allow B-balls to form \cite{qb1}. However,  
since the energy per 
unit charge in this case is roughly independent of $B$, the B-balls are 
unstable with respect to
decay to quarks or nucleons. 
In a previous paper \cite{qb1} we noted that the B-balls can decay at temperatures less 
than
 that of the electroweak phase transition, $T_{ew}$, and that
 the decay of such B-balls could have important implications 
for baryogenesis. For example, they could protect the B asymmetry from the effects of L violating interactions
above $T_{ew}$, when anomalous $B+L$ violation is in thermal equilibrium, 
or allow a baryon asymmetry
to come from a $B-L$ conserving condensate \cite{qb1}. One goal of this paper 
is to consider in more detail the physics
of B-ball formation, thermalization and decay. 

     When the B-balls decay at temperatures below $T_{ew}$, 
the observed baryon number will be a combination of the 
baryon number originating from the decay of the B-balls and that from the 
free squarks left over
after the break up of the squark condensate. 
Since the B-balls are composed of squarks, when they
decay they will naturally produce a $number$ density of neutralinos of the same 
order of magnitude as
the number density of baryons. 
Therefore, if the B-balls decay sufficiently below the freeze-out temperature 
of the 
lightest supersymmetric particle (LSP) neutralinos and if the number
 density of thermal relic neutralinos is less than that from
B-ball decay, then the dark matter density and baryon density in the Universe will 
be naturally related by the similarity of their number densities.

		  The actual ratio of baryons to dark matter will essentially be determined by two variables: (i) the mass of the neutralino LSP
 and (ii) the proportion of baryon number trapped in B-balls to that in free squarks after the break-up of the AD condensate (which we will 
refer to as the efficiency of B-ball formation). 
 We will attempt to show that the baryon to dark matter ratio expected in 
BBB is typically of the form observed in the Universe, i.e. 
that it is naturally less than 1 but not very much smaller than 1.    
 
		 We first give an expression for the baryon to dark
 matter ratio in BBB. A B-ball will contain a B asymmetry in the form of a squark
 field. When the B-ball decays, for each unit of B produced, corresponding to the decay of 3
 squarks to quarks, there will be at least three units of R-parity produced, corresponding to at least 3 neutralino LSP's. 
(Depending on the nature of the cascade produced by the squark decay
 and the LSP mass, more LSP pairs could be produced). Let $N_{\chi} \gae 3$ be the number of
 LSPs produced per baryon number. Let $f_{B}$ be the fraction of the total B asymmetry which is 
contained in the B-balls. Then the baryon to dark matter ratio will be given by 
\be{i1} r_{B} = \frac{\rho_{B}}{\rho_{DM}} = \frac{m_{n}}{N_{\chi} f_{B} m_{\chi}}  ~,\ee
 where $m_{n}$ is the nucleon mass and $m_{\chi}$ is the neutralino LSP mass. 
We first suggest that it is rather natural to have $r_{B} < 1$. The present LEP lower bound on the 
neutralino mass in the MSSM, assuming no constraints on the scalar masses, is 17
GeV \cite{olivedm}. If we 
were to assume radiative electroweak symmetry breaking and universal masses for the squarks
 and Higgs scalars at the unification scale, then the lower bound would become 
$m_{\chi} \gae 40\GeV$ for $tan \beta \lae 3$ \cite{olivedm}. For $N_{\chi} \geq 3$, and 
with $m_{\chi} \gae 17 \: (40)\GeV$,
 we find that $r_{B} < 1$ occurs for $f_{B} \gae 0.02 \;(0.008)$. Thus so long as more than 
2$\%$ of the baryon asymmetry is trapped in B-balls, the observed dominance of 
dark matter in the Universe will be naturally explained. We will show that this is 
likely to occur.

	Primordial nucleosynthesis \cite{sarkar,eu} bounds the density of baryons in the Universe to satisfy 
$ 0.0048 \lae \Omega_{B}h^{2} \lae 0.013$, where $0.4 < h < 1$. (We adopt the bound 
based on "reasonable" limits on primordial element abundances \cite{sarkar}). Thus 
the observed baryon to dark matter ratio, $r_{B} \approx  
\Omega_{B}/(1-\Omega_{B})$
(assuming a flat Universe),
 satisfies $0.005 \lae r_{B} \lae 0.09$. This can be accounted 
for by BBB if
\be{i2} 3.7\GeV \lae \left(\frac{N_{\chi}}{3}\right) f_{B}m_{\chi} 
\lae 67\GeV  ~.\ee
For example, if the LSP mass satisfies $17 \; (40)\GeV \lae m_{\chi} \lae 500\GeV$, 
then the 
observed baryon to dark matter ratio can be satisfied by a wide range of 
$f_{B}$, 
$0.007 \lae f_{B} (N_{\chi}/3)$ 
$ \lae 3.9 \; (1.7)$. We also note that if 
$m_{\chi} \gae 67\GeV$ then we 
must have $f_{B} < 1$, implying that the observed baryon 
asymmetry must come from a $mixture$ 
of decaying B-balls and free baryons. 

	     This all assumes that the asymmetry not trapped in the B-balls can survive down to 
temperatures below $T_{ew}$. However, if we were to consider a $B-L$ conserving condensate
or additional $L$ violating interactions in thermal equilibrium above $T_{ew}$, then the only B
asymmetry which could survive anomalous B+L violation is that
 associated with the B-balls. In this case, $f_{B}$ would be effectively
 equal to 1 (we refer to this case as "pure" BBB) 
and so $m_{\chi}$ would have to be less than 67$\GeV$. 

	A crucial assumption in all this is that there is effectively no subsequent annihilation of the 
LSP's coming from B-ball decays. If this were not the case, then
we would lose the similarity of the number densities of baryons and 
dark matter particles which is essential if we are to explain
the baryon to dark matter ratio naturally via B-ball decays.

	     In order to find out if BBB can naturally
 account for the baryon to dark matter
ratio in the MSSM, we must consider in some detail the formation of B-balls from the primordial AD
 condensate, their survival in the thermal environment following reheating and their eventual decay to
 baryons and dark matter. The paper is organized as follows. 
In section 2 we discuss the efficiency of B-ball formation in a minimal
 cosmological scenario. In section 3 we consider the condition for the B-balls to survive thermalization
 down to temperatures less than that of the electroweak phase transition. In section 4 we
 discuss the decay of the B-balls and the resulting LSP dark matter density. In section 5 we apply our results to realistic examples. 
In section 6 we give our conclusions. In the Appendix we discuss some aspects
 of B-ball solutions of the scalar field equations.

\section{B-ball formation}                                           

We will assume that the B asymmetry is conserved from the time of 
AD condensate formation until the present.
In order for the baryon to dark matter ratio to be explained by B-ball decay
in the MSSM it is necessary that B-balls can be formed efficiently. This means 
that greater than around 2$\%$ of the B asymmetry
 present in the original condensate should be trapped in the B-balls. 
B-balls form because of the attractive force due to the logarithmic radiative correction term in
the condensate scalar potential \cite{qb1}
\be{e1} U(\Phi) \approx m^{2}\left(1 +  K log\left( \frac{|\Phi|^{2}}{M^{2}} \right) \right) |\Phi|^{2} 
+ \frac{\lambda^{2}|\Phi|^{2(d-1)}
}{M_{p}^{2(d-3)}} + \left( \frac{A_{\lambda} 
\lambda \Phi^{d}}{d M_{p}^{d-3}} + h.c.\right)    ~,\ee
where $d$ is the dimension of the non-renormalizible term in the superpotential. Of particular interest to us here will be the 
d=6 $\udd$ "squark" direction, with a non-renormalizible superpotential term of the form $(\udd)^{2}$,
 and the d=4 $\uude$ direction, which conserves $B-L$. (F- and D-flat directions in the MSSM are classified in 
reference \cite{drt}).
The logarithmic term causes any space-dependent perturbation in the condensate field to grow and go
non-linear at some time, leading to the formation of B-balls.

	 The magnitude of $K$ will be important in our numerical estimates. From the 1-loop 
effective potential \cite{nilles}, for the $\udd$ direction, the correction
 due to gauginos with SUSY breaking masses $M_{\alpha}$ is given by 
\be{e1a} K \approx -\frac{1}{3} \sum_{\alpha, \; gauginos} 
\frac{\alpha_{g_{\alpha}}}{8 \pi} \frac{M_{\alpha}^{2}}{m^{2}}  
  ~,\ee
where the sum is over those gauginos which gain a mass from the condensate scalar $\phi$. 
The main contribution will come from the three gluinos which gain masses from the 
squark expectation values. With $\alpha_{g_{3}} \approx 0.1$ we obtain 
$|K| \approx 0.004 (M_{3}/m)^{2}$. Thus, depending on the ratio of the SUSY breaking
gluino mass to the squark mass, we expect $|K|$ to be typically in the range 0.01 to 0.1.

\subsection{Cosmological scenario}

		  In one particularly likely cosmological scenario, the AD scalar obtains a negative order $H^{2}$ 
correction to its mass squared term \cite{drt,h2}, $m^{2} \rightarrow (m^{2} - c H^{2})$  with $c \approx 1$,  
as a result of non-minimal kinetic terms coupling the AD scalar to the inflaton and other fields in the supergravity K\"ahler potential.
After inflation, the energy density of the Universe will be matter dominated by the oscillations of the inflaton field
about the minimum of its potential. This will continue until the inflaton completely decays, 
leaving a radiation dominated Universe at the reheating temperature $T_{R}$,
 where $T_{R}$ must be less than 
about $10^{8-9}\GeV$ for $m_{3/2}$ in the range 100 GeV to 1 TeV in order to avoid thermally regenerating gravitinos \cite{sarkar}. 
In this scenario there is an initial spectrum of perturbations of the magnitude of AD field, 
due to quantum fluctuations during inflation, given by \cite{qb1}
\be{e2} \delta \phi_{o}(\lambda_{o}) \approx
 \frac{1}{2 \pi m_{S} H_{I}^{1/2} \lambda_{o}^{5/2}}    ~.\ee
Here $H_{I}$ is the Hubble parameter during inflation and $\lambda_{o}$ is the length scale at 
$H^{2} \approx m^{2}$. 
This is the spectrum of perturbations when $H^{2} \approx m^{2}$, at which
 time the effective mass squared term
changes sign and the AD field starts oscillating, forming the condensate.
(In the following we will use the convention that
subscript $o$ denotes the time at which the AD condensate starts oscillating and 
$i$ denotes the time when the perturbations forming the B-balls go non-linear). 
 
\subsection{Growth of perturbations}

	    In order to discuss the linear evolution of the perturbations,
 it is convenient to consider the special case of
a homogeneous condensate of the form \cite{ks,lee}
\be{e3} \Phi = \frac{\phi(t)}{\sqrt{2}}e^{i \theta(t)}       ~,\ee
 where $\phi(t) = (a_{o}/a)^{3/2}\phi_{o}$ and
$\dot{\theta}(t)^{2} \approx m^{2}$, up to corrections of order $Km^{2}$, and 
where we assume that $|K|$ is small compared with 1.
This describes a condensate with the maximum possible charge asymmetry for a
 given maximum amplitude of $\Phi$. Since it may be shown that 
the growth of perturbations during the 
linear regime is unaffected by the charge of the condensate, the 
results obtained for this homogeneous condensate
 (which is particularly easy to perturb about) 
will also apply to the 
more general case with a smaller charge asymmetry. 

 The linearized equations for the perturbations 
$\phi = \phi(t) + \delta \phi(x,t)$ and $\theta = \theta(t) + \delta \theta (x,t)$
are given by \cite{ks}
\be{e4} \delta \ddot{\phi} + 3H \delta \dot{\phi} -(2 \dot{\theta}(t) \phi(t) \delta \dot{\theta} + 
\delta \phi \dot{\theta}(t)^{2}) -\frac{\nabla^{2}}{a^{2}} \delta \phi 
= - U^{''}(\phi(t)) \delta \phi     ~\ee 
and
\be{e5} \phi(t) \delta \ddot{\theta} + 
3H (\dot{\theta}(t) \delta \phi + \delta{\dot{\theta}}\phi(t)) + 
2( \dot{\phi(t)} \delta \dot{\theta} + \delta \dot{\phi} \dot{\theta}(t))
- \frac{\phi(t)}{a^{2}} \nabla^{2} \delta \theta = 0   ~.\ee
Assuming a solution of the form \cite{ks} 
\bea{e6} \delta \phi &=& \left(\frac{a_{o}}{a}\right)^{3/2} 
\delta \phi_{o} e^{i(S(t) + \vec{k}\cdot
\vec{x})}     ~\nn
\delta \theta &=&  \delta \theta_{o} e^{i(S(t) + \vec{k}\cdot
\vec{x})}     ~\eea
where $\delta \phi_{o}$ and $\delta \theta_{o}$ are the initial values of the perturbations,
and, solving for $S(t)$, the perturbations grow according to
\be{e8} \delta \phi = \left(\frac{a_{o}}{a}\right)^{3/2}
 \delta \phi_{o}\;  exp \left( \int dt \left(\frac{1}{2} \frac{\vec{k}^{2}}{a^{2}}
 \frac{|K| m^{2}}{\dot{\theta}(t)^{2}} \right)^{1/2} \right)  
e^{i\vec{k}.
\vec{x}}   
  ~\ee
and 
 \be{e9} \delta \theta \approx \delta \theta_{i}\;  exp \left( \int dt \left(\frac{1}{2} \frac{\vec{k}^{2}}{a^{2}}
 \frac{|K| m^{2}}{\dot{\theta}(t)^{2}} \right)^{1/2} \right)   
e^{i\vec{k}\cdot
\vec{x}}       ~.\ee
These apply if 
$ \left| \vec{k}^{2}/a^{2} \right| \lae | 2K m^{2}|   $,              
$H^{2}$ is small compared with $m^{2}$ and
$|K| \ll 1$. 
If the first condition is not satisfied, the gradient energy of the perturbations produces a positive
 pressure larger than the negative pressure due to the attractive force from the logarithmic term, 
preventing the growth of the perturbations.

\subsection{Evolution of the condensate lumps}

	       The initial size and charge of the non-linear lumps of condensate 
 will be determined by the scale
of the first perturbation to go non-linear.
When the lumps initially go non-linear, they will be roughly spherically symmetric. We 
will model these lumps by a homogeneous condensate field of the form
\bea{e10} \Phi &=& A(t) Cos (mt) + i B(t) Sin (mt) \;\;\;\; ; \;\;\; r \leq R(t)  ~\nn
\Phi &=&  0  \;\;\;\; ; \;\;\; r > R(t)  ~,\eea
where $A(t) \gg B(t)$. (From now on we will consider a condensate with a small 
charge asymmetry). 
 Non-zero $B(t)$ gives the condensate a charge,
\be{e12} Q = - i \int d^{3}x (\Phi^{\dagger} \dot{\Phi} - \dot{\Phi}^{\dagger} \Phi)
 = \frac{8 \pi}{3}R^{3}mAB   ~,\ee
where in the final expression we have assumed that $\dot{A}B = A\dot{B}$, which will 
be shown to be consistent. 

	       Physically, we can understand how the lump will evolve. Roughly speaking, the particles in the
 condensate lump, which we take to be initially at rest, will be attracted towards each other by the force from the 
logarithmic term. Thus the radius of the condensate lump will decrease, and the value of $A(t)$ will increase. 
This will continue until the value of $A$ becomes so large that the non-renormalizible term in the 
scalar potential or the gradient energy dominates the logarithmic term, 
at which point the force between the particles in the lump 
will become rapidly repulsive and the lump will "bounce" elastically and start to expand. In the
 absence of a mechanism to dissipate the energy, the lump would simply pulsate. We consider first
the condition for this pulsating lump to be able to retain a given charge. Assuming that the 
lump remains homogeneous throughout (there could be formation of a
 thin-wall of width of the order of $m^{-1}$, giving the lump the 
profile of a thin-wall bubble, but this will 
not affect the dynamics significantly), 
the energy of the lump, up to terms proportional to $|K|$, 
 will be given by $E \approx E_{l} + E_{q} $, 
where 
\be{e13} E = \int d^{3}x \; |\dot{\Phi}|^{2} +
 |{\nabla}\Phi |^{2}  + U(\Phi)  ~,\ee
is the total energy of the condensate lump,
\be{e14} E_{l} \approx \frac{4 \pi}{3} R^{3} m^{2} A^{2}  ~\ee
is the energy in of lump in the absence of charge and
\be{e15} E_{q} \approx \frac{4 \pi}{3} R^{3} m^{2} B^{2}  ~\ee
is the energy associated with the charge.
Since the energy of the lump is constant as $R$ decreases, we have $A(t) \propto R^{-3/2}$. 
If the charge is trapped in the lump and so is conserved, we will have 
$B(t) \propto 1/(AR^{3})  \propto R^{-3/2}$. (This implies that $\dot{A}B = A\dot{B}$, as we
have assumed). Thus the energy associated with the
charge ($\propto R^{3}B$) will remain constant as the lump contracts. Therefore if the energy per unit charge
 of the initial lump is
small enough for the charge to be trapped in the lump, $E_{q} < mQ$, then it will remain trapped 
throughout the pulsation of the lump. This condition is satisfied if 
\be{e16} Q \lae Q_{max} = \frac{16 \pi}{3} m R^{3} A^{2} ~.\ee
Thus if we consider the initial condensate to have a charge asymmetry density corresponding to the 
baryon asymmetry of the Universe, $n_{B}$, we see that the
 whole charge within the initial lump can be trapped if initially 
$n_{B\;i} \lae 4 m A_{i}^{2}$, where $A_{i}$ is the initial amplitude of the scalar field and $n_{B\;i}$ is the baryon asymmetry when the lump first goes non-linear. 

             In general, the energy of the initial lump will be larger than that of the final B-ball of the same
charge. Thus there must be some mechanism by which the lump can lose energy. For example, this
 could occur by the radiation of classical scalar field waves
 created by the pulsation of the lump. We will not attempt to 
discuss the details of the energy loss mechanism here, since our arguement
 is based purely on the fact that it is energetically favourable for the charge to remain within the lump. 
The result of such an energy loss mechanism is that the 
value of $R$ and $A$ when the lump reaches its largest size will decrease with time. 
However, by charge conservation, since the charge is proportional to $ABR^{3}$, the value of $B$ will increase. Eventually the values of 
$A$ and $B$ will become similar, at which point the charge of the lump will dominate its 
subsequent evolution.
The B-ball will form soon after this point is reached. The $smallest$ possible values of $R$ and $A$ during the evolution of the condensate lump will correspond to the values 
 associated with a B-ball of the corresponding
 charge, $R_{B}$ and $A_{B}$. Thus so long as the condition $B \lae B_{max}(A_{B}, R_{B})$ is still satisfied,
 we expect that the B-ball will be able to form from the collapsing condensate 
lump without any loss of charge. 

	      In general, a B-ball can either be thick or thin-walled. 
In the case of the thin-walled B-ball, the radius is a function of charge
whilst the value of $A$ remains constant 
\be{e17} A_{B} = A_{c}\; ; \;\;\; R_{B} = \zeta B^{1/3}  ~,\ee
where $A_{c}$ is the value of the field inside the thin-wall B-ball
(we discuss some properties of B-ball 
solutions in the Appendix).
A thin-walled B-ball will form if the charge is greater than some critical charge, $B_{c}$, which depends on 
details of the scalar potential, in particular on its dimension $d$. 
In the case of the thick-walled B-ball, corresponding to a charge less than $B_{c}$, 
the radius remains approximately constant whilst $A$ becomes charge dependent
\be{e18} A_{B} \approx \eta B^{1/2} \; ; \;\;\;  R_{B}\approx R_{c} ~.\ee
In both cases, we see that the condition for a lump of a given charge B to be able to
form a B-ball without losing charge, $B \lae B_{max}(A_{B}, R_{B})$, reduces to
a B-independent condition (which we will refer to as the B-ball condition); for the thin-walled case, 
\be{e19} 1 \lae \frac{16 \pi}{3} m \zeta^{3} A_{c}^{2}   ~\ee
whilst for the thick-walled case,
\be{e20} 1 \lae \frac{16 \pi}{3} m R_{c}^{3} \eta^{2}   ~.\ee
Therefore we can conclude that if the charge of the initial lump satisfies the 
condition to be trapped in the lump ($n_{B} \lae 4 m A_{i}^{2}$) and 
if the appropriate B-ball condition is satisfied, then 
B-balls may be expected to form with a high efficiency, with a large percentage of the 
charge of the condensate being trapped in the B-ball. 

	   In fact, we can show that the B-ball conditions will be generally satisfied for any B-ball. 
We first review some properties of thin and thick-walled B-balls (see Appendix). 
In general the B-ball is described by \cite{cole}
\be{e21} \Phi(r,t) = \frac{\phi(r)}{\sqrt{2}} e^{i \omega t}     ~,\ee
where $\phi(r)$ and the constant 
$\omega$ are found by minimizing the energy of the scalar field configuration for a fixed charge. 
For the thin-walled B-ball, the energy per unit charge and radius are given by \cite{cole}
\be{e22} \frac{E}{B} = \left(\frac{2U(\phi_{c})}{\phi_{c}^{2}}\right)^{1/2}    ~\ee
and
\be{e23} R = \left(\frac{3}{4 \pi}
 \frac{B}{\left(2 \phi_{c}^{2} U(\phi_{c}) \right)^{1/2}}\right)^{1/3}     ~.\ee
From numerically solving for the B-ball for $|K|$ in the range 0.01 to 0.1,
 we find that the thick-wall B-ball radius is given by 
$R_{c} \approx k_{R} (|K|^{1/2}m)^{-1}$, where $k_{R} \approx 1.2$ (see Appendix).
 (From now on we will use $R_{c} \approx (|K|^{1/2}m)^{-1}$). The critical charge
$B_{c}$ corresponds to  the charge at which the thin-wall radius,
\eq{e23}, is approximately equal to $R_{c}$,
\be{e24} B_{c} \approx  \left( \frac{4 \pi}{3} \right)
 \frac{\left(2 \phi_{c}^{2} U(\phi_{c}) \right)^{1/2}}{|K|^{3/2} m^{3}}     ~.\ee
The field inside the thin-walled B-ball, $\phi_{c}$, corresponds to the minimum of $E/B$ as a function 
of $\phi$. 
The value of $\zeta$ is then set by the values of $R$ and $B$ at which
 the thin-wall B-ball reaches the thick-wall limit, 
\be{e25} \zeta \approx \left( \frac{3}{4 \pi} \right)^{1/3}
 \frac{1}{\left(2 \phi_{c}^{2} U(\phi_{c}) \right)^{1/6}}        ~.\ee
The thin-walled B-ball condition then becomes
\be{e26}   1 \lae \frac{2 mB}{{E}}     ~,\ee
where we have used $A_{B}^{2} =  A_{c}^{2} \equiv \phi_{c}^{2}/2 $. Since for the B-ball
the energy per unit charge, $E/B$, is always less than $m$, this condition will be generally satisfied. 
Similarly, for the thick-walled B-ball, we can fix the value of $\eta$ when the charge equals 
$B_{c}$ and the field at the centre of the thick-walled B-ball equals $A_{c}$. 
The thick-walled B-ball condition is then found to be the same as for the
thin-walled case. Thus we can conclude that the B-ball condition will be generally satisfied. 

\subsection{B-ball charge}

	       The initial charge of the B-balls will be determined by the wavelength of 
the first perturbation mode to go non-linear. This will correspond to the first perturbation
which can result in $\delta \phi/\phi \approx 1$ in a time of the order of $H^{-1}(t)$. For an inflaton
oscillation matter dominated Universe, the exponential growth factor is
\be{g1} \int dt \left(\frac{1}{2} \frac{\vec{k}^{2}}{a^{2}}
 \frac{|K| m^{2}}{\dot{\theta}(t)^{2}} \right)^{1/2} = \frac{2}{H} \left(
\frac{|K|}{2} \frac{\vec{k}^{2}}{a^{2}}\right)^{1/2}      ~.\ee 
(We take the scale factor when the AD oscillations begin to be equal to 1).
The largest growth factor will correspond to the largest value of $\vec{k}^{2}$ for which growth
can occur, $\vec{k}^{2}/a^{2} \approx 2 |K| m^{2}$. Thus the value of $H$ at which the
first perturbation goes non-linear will be
\be{g2} H_{i} \approx \frac{2 |K| m}{\alpha(\lambda_{o})}      ~,\ee
with
\be{g3} \alpha(\lambda_{o}) = - log \left( \frac{\delta \phi_{o}(\lambda_{o})}{\phi_{o}}\right)    ~,\ee
where $\lambda_{o}$ is the length scale of the perturbation at $H \approx m$
and $\phi_{o}$ is the value of $\phi$ when the AD condensate oscillations begin.
The charge of the condensate lump is determined by the baryon asymmetry of the Universe at $H_{i}$ 
and the initial size of the perturbation when it goes non-linear. The initial non-linear region has a 
radius $\lambda_{i}$ at $H_{i}$ given by
\be{g4} \lambda_{i} = \frac{\pi}{(\alpha(\lambda_{o}) m H_{i})^{1/2}}    ~.\ee
The baryon asymmetry of the Universe at a given value of $H$ during inflaton oscillation
domination is given by 
\be{g5} n_{B} = \left( \frac{\eta_{B}}{2 \pi} \right) \frac{H^{2} M_{Pl}^{2}}{T_{R}}  \approx 
1.6 \times 10^{18} H^{2} \left(\frac{10^{9}}{T_{R}}\right)   
\left(\frac{\eta_{B}}{10^{-10}}\right)   ~,\ee
where we have taken the baryon to entropy ratio to be $\eta_{B} \approx 10^{-10}$. 
 Thus the charge in the initial condensate lump is given by 
\be{g6} B = \frac{4 \pi^{3}}{3 \sqrt{2}} 
\frac{\eta_{B} |K|^{1/2} M_{pl}^{2}}{m \alpha^{2} T_{R}} 
= 2 \times10^{15} |K|^{1/2} 
\left(\frac{100 \GeV}{m}\right)
\left(\frac{10^{9} \GeV}{T_{R}}\right)
\left(\frac{40}{\alpha}\right)^{2}
\left(\frac{\eta_{B}}{10^{-10}}\right) 
~,\ee 
where we have used $\alpha(\lambda_{o}) = 40$ as a typical value.
It is important to note that the details of the initial spectrum of perturbations and the 
condensate potential only enter logarithmically through $\alpha(\lambda_{o})$. 
The condition for the B-balls to form efficiently, \eq{e16}, may then be written as 
\be{g7}  T_{R} \gae \frac{\eta_{B} m M_{Pl}^{2}}{8 \pi A_{o}^{2}}
= 0.23 \left(\frac{m}{100 \GeV}\right)
\left(\frac{4.1 \times 10^{14} \GeV}{A_{o}}\right)^{2} 
\left(\frac{\eta_B}{10^{-10}}\right)\GeV   ~,\ee
where $A_{o}= 4.1 \times 10^{14}\GeV$ is a typical initial value for 
the d=6 $\udd$ direction. 
We will apply these results to realistic examples later. 

\section{B-ball thermalization}

	     We next consider the conditions under which the B-balls can 
survive down to 
temperatures less than that of the electroweak phase transition $T_{ew}$. 
In order to discuss this question, we will consider throughout a Gaussian 
Ansatz for the thick-walled 
B-ball
\be{t1} \phi(r) = \phi(0) e^{-\frac{r^{2}}{R^{2}}}    ~,\ee
where $\phi(0)$ is the value of $\phi$ at the center of the B-ball.
 In the Appendix it is shown that the 
Gaussian is a physically reasonable approximation for the thick-walled B-ball. 

     There are two processes which might lead to the destruction 
of B-balls by the thermal 
background. The first process is the dissociation of the B-ball 
by collisions of thermal particles with
 the "hard core" of the B-ball, corresponding the region of the 
B-ball that thermal particles cannot penetrate. (This was considered first in reference \cite{ks}).
 The second process is the dissolution of the B-ball by the thermalization
 and transport of charge from the "soft edge"
 of the B-ball, within which a thermal equilibrium distribution of 
background particles can exist \cite{qb1}. 

\subsection{Dissociation}

		 We first consider the dissociation of the B-ball. Thermal particles can penetrate the 
B-ball to the radius at which $g\phi(r) \approx 3 T$; we refer to this as the stopping radius $r_{st}$, 
with the corresponding field being $\phi_{st}$. (This defines the hard core of the B-ball).
 As the thermal particles enter the B-ball, they will
transfer energy to the classical scalar field, finally coming to a (temporary) halt. If the time scale
over which the particle stops is short compared with the time over which the scalar field of the B-ball can
absorb the energy, $\delta t_{e} \sim 1/m$ (set by the dynamical scale of the scalar field),
 then the thermal particle will be 
ejected, leaving some energy in the hard part of the B-ball, with the amount of energy
 transferred depending on how quickly the particle comes to a halt. 
Let $\delta t_{r} = K_{r}/m$ ($K_{r} \gae 1$) be the time scale over which an excited
 B-ball can radiate its excess energy in the form of scalar field waves.
If sufficient energy can be delivered to the B-ball within
$\delta t_{r}$ to overcome the binding energy of the charges in the B-ball, then the B-ball
will dissociate. 
If, on the other hand, less energy is delivered in this time, then the 
B-ball will be able to radiate the excess energy adiabatically and so will not dissociate. 
Thus the condition for the B-ball to evade dissociation is that $ \Delta E(\delta t_{r}) 
< \delta m B$, where $\Delta E(\delta t_{r})$ is the energy delivered to the B-ball
by thermal particles in a time $\delta t_{r}$ and $\delta m = (m-E/B) \approx |K|m$ (see Appendix) 
is the binding energy per unit charge.

	   The flux of particles onto the hard core of the B-ball is 
\be{t1} f = \frac{\tilde{g}(T)}{\pi^{2}} 4 \pi r_{st}^{2} T^{3}     ~\ee
 where $\tilde{g}(T) \approx 100$ is the effective number of light 
thermal degrees of freedom
coupling to the condensate scalars, 
and, for the Gaussian thick wall B-ball,
\be{t2} r_{st} = \beta R \;;\;\;\;\; \beta = 
log^{1/2}\left(\frac{g \phi(0)}{3 T}\right) \;\; , \;\; \beta > 1   ~.\ee
The energy transferred in a collision will depend on how quickly the 
$\phi$ field of the 
B-ball can respond to the impacting thermal particle. The time scale 
over which a 
particle of energy $3 T$ comes to a halt in the hard core is roughly 
the distance over which 
the change in the effective mass of the incoming particle, $ g \delta\phi$, 
is approximately equal to $3 T$ around $r_{st}$. For the Gaussian
wall this corresponds to
\be{t3} \delta r_{st} \approx \frac{R^{2}}{2 r_{st}} = \frac{R}{2 \beta}     ~.\ee
Thus we expect the 
suppression factor in the transfer of energy to be of the order of 
$m^{-1}/\delta r_{st} = 2 \beta/(mR) \approx 2 \beta |K|^{1/2}$, which
is typically less than 1 but not very much so. Therefore we do not 
expect a very large suppression in the energy transferred. 
(In fact, we will show later that a thermal equilibrium exists within the soft edge of the B-ball.
 Therefore thermal scattering may slow an incoming particle,
 so that the B-ball will have time to absorb most of its energy).
Let the energy per thermal particle transferred to the B-ball be 
$\gamma_{T} T$, where $\gamma_{T} \lae 3$.  
Then the rate of energy increase due to incoming thermal particles will be 
\be{t4}  \frac{dE}{dt} = \frac{4 \tilde{g}(T) \gamma_{T} T^{4} 
\beta^{2} R^{2}}{\pi}     ~.\ee
The requirement to avoid dissociation is that, in the time $\delta t_{r}$, 
insufficient energy is 
supplied to the B-ball to overcome the binding energy. This gives an 
upper bound on the temperature at which B-balls can exist:
\be{t5} T \lae \left[\frac{\pi |K|}{4 \tilde{g}(T) K_{r} \gamma_{T} 
\beta^{2}}
 \left(\frac{\delta m}{m}\right) \right]^{1/4} m B^{1/4}
    ~.\ee
Combining this with the expression for the B-ball
charge as a function of the reheating temperature, \eq{g6}, then gives
an upper bound on the reheating temperature
\be{t6} T_{R} \lae 1 \times 10^{6} |K|^{3/10} 
\left(\frac{m}{100\GeV}\right)^{3/5}
\left[ 
\frac{1}{K_{r} \gamma_{T} \beta^{2}}
 \left(\frac{\delta m}{m}\right) 
\right]^{1/5} 
\left(\frac{40}{\alpha}\right)^{2/5} \left(\frac{\eta_B}{10^{-10}}\right)^{1/5}\GeV
~,\ee  
where we have taken $\tilde{g}(T) \approx 100$.
Thus, using as typical values $K_{r} \approx 10$, $\alpha \approx 40$, 
$\beta \approx 4$, $\delta m/m \approx |K|$ and 
$\gamma_{T} \approx 1$, we obtain for $|K| \approx 0.01\;(0.1)$
\be{t6a} T_{R} \lae 3\;(9) \times 10^{4} 
\left(\frac{m}{100\GeV}\right)^{3/5} \GeV   ~.\ee
Therefore, so long as $T_{R} \lae 10^{4-5}\GeV$, the B-balls should be able 
to evade dissociation for $|K|$ in the range 0.01 to 0.1. 

\subsection{Dissolution}

	   The second process which might lead to the thermalization
 of the B-ball is dissolution by the removal of charge from the soft edge of the B-ball. 
The width of the soft edge will correspond to the distance at $r_{st}$ over which $\phi$ 
does not change much, $\delta \phi/\phi \lae 1$. For the Gaussian B-ball this corresponds to
 $\delta r_{st} \approx R/(2 \beta) \approx m/(2 |K|^{1/2}\beta)$. This is much larger
than the mean free path of the strongly interacting thermal
 background quarks, $\lambda_{mfp} \approx k_{q}/T$ (where $k_{q} 
\approx 6$ \cite{qmfp}), if $T \gae m$. 
So we expect a thermal equilibrium to exist within the soft edge at 
high temperatures.
 We also expect the B-ball charges within the soft edge to become thermalized. 
The rate at which charge is  removed from the B-ball will
then depend on two factors. Firstly, how rapidly the thermalized 
charges diffuse out of the B-ball.
Secondly, how rapidly the hard core of the B-ball reconfigures itself 
in order to compensate for the loss
of charge and so minimize its energy, replenishing the charge in the 
soft edge. 
The rate of diffusion may be estimated as follows. The number of steps for 
a charge to leave the soft edge
by a random walk will be of the order of $(\delta r_{st}/
\lambda_{mfp})^{2}
 \approx (RT/(2 k_{q} \beta))^{2}$.
 The time for each step is around $\lambda_{mfp}$.
 Therefore the time taken for a charge to leave the soft edge is 
\be{t7} \tau_{d} \approx \left(\frac{R}{2 \beta}\right)^{2}
\frac{T}{k_{q}}     ~.\ee
The total charge in the soft edge is 
$B_{soft} \approx 4 \pi \omega \phi_{st}^{2} r_{st}^{2} \delta r_{st}$.
 Thus the 
rate at which charge can be removed, assuming that the B-ball can 
reconfigure itself on a time scale shorter that
$\tau_{d}$, is 
\be{t8} \frac{1}{B} \frac{dB}{dt} \approx \frac{1}{B} 
\frac{B_{soft}}{\tau_{d}} ~,\ee 
where $B \approx 2 \omega \phi(0)^{2}
R^{3}$ is the total charge of the Gaussian B-ball.
 The time for the B-ball to reconfigure  itself will be
of the order of $m^{-1}$,
 which will be short compared with $\tau_{d}$ for $T \gae m_{*} = 4 \beta^{2} k_{q} |K| m$. 
So at high temperatures we expect that the B-ball can efficiently 
replenish the charge lost by thermalization of the
edge. The rate at which charge is lost is then 
\be{t9} \frac{1}{B} \frac{dB}{dt} \approx \frac{4 \pi  \beta^{3} k_{q}
T}{g^{2} \phi(0)^{2} R^{2}}    ~.\ee 
Requiring that this is less than
$H$, in order to avoid dissolution of the B-ball, then imposes a lower
limit on the B-ball charge
\be{t10} B \gae \frac{8 \pi \omega k_{q} \beta^{3} R}{g^{2}  k_{T}}
\frac{M_{Pl}}{T}    ~,\ee where $H = k_{T}T^{2}/M_{Pl}$ ($k_{T} \approx
17$).  For $T \lae m_{*}$, the rate at which charge is lost will be
determined by the rate at which the B-ball can replenish the lost
charge, in which case 
$T$ in \eq{t10} should be replaced by $ m_{*} $ and the lower bound on $B$ becomes $T$ independent. 
Requiring that \eq{t10} is satisfied for all $T \gae m_{*}$ then gives the lower bound on $B$,
\be{t11} B \gae 1.4 \times 10^{17} \;
\frac{1}{g^{2} |K|^{3/2}}
\left( \frac{\beta}{4} \right) 
\left( \frac{100\GeV}{m} \right) 
~.\ee 
Using \eq{g6} we then obtain an upper bound
on the reheating temperature 
\be{t12} T_{R} \lae 1.3 \times 10^{7} \;
g^{2} |K|^{2}              
\left( \frac{4}{\beta} \right) 
\left( \frac{40}{\alpha} \right)^{2}
\left(\frac{\eta_B}{10^{-10}}\right)\GeV ~.\ee
Therefore, with $g \approx 1$, $k_{q} \approx 6$ and $|K| = 0.01 \;(0.1)$, we find that
 the reheating temperature
cannot be larger than $10^{3} \; (10^{5})\GeV$ if the B-balls are to have sufficiently 
large charge to evade dissolution. 
We note that this bound is tighter than that from dissociation for all $|K| \gae 0.01$. 

	      Thus we can conclude that dissociation of B-balls by thermal particle collisions
 and dissolution by soft edge thermalization
can both be avoided if the reheating temperature is sufficiently low, 
less that about $10^{3}-10^{5}\GeV$ for $|K|$ in the range 0.01 to 0.1. 

\subsection{Dissociation prior to reheating}

	The above discussion applies to the radiation dominated period following reheating. The 
"reheating" temperature in fact refers to the temperature at which the radiation from the 
decay of the inflaton condensate comes to dominate the matter density remaining in the condensate. 
The temperature decreases steadily during the inflaton oscillation dominated period, with no 
"reheating" as such. The  
radiation energy density due to inflaton decays is given by \cite{eu,drt}
\be{t13} \rho_{r} \approx \frac{2 \Gamma \rho}{5 H}     ~,\ee
 where $\Gamma$ is the decay rate of the inflaton condensate particles, which  have a 
mass density $\rho$. 
Therefore we must ensure that this radiation energy density 
does not dissociate the B-balls prior to reheating. 
(Dissolution by thermalization of the soft edge of the B-ball is most 
effective at $lower$ temperatures, since the rate of dissolution, \eq{t9}, is proportional to $T$, whereas $H$ 
is proportional to $T^{2}$ for radiation domination and
 $T^{4}$ for inflaton domination. Therefore the possibility
 of higher temperatures existing prior to reheating will not alter the upper bound on $T_{R}$ following from dissolution).
In addition to the inflaton condensate, there is a second source of radiation, namely the remains of the original squark 
condensate. After the AD condensate breaks up into B-balls and non-relativistic
squarks (the initial momentum of the squarks radiating from the collapsing condensate lumps 
will be at most of the order of the inverse of the 
radius of the B-ball, $|K|^{1/2}m$), the non-relativistic squarks will still evolve as a matter density much like a condensate. 
The average value of the scalar field associated with this density of squarks will be of the same 
order as the original AD field amplitude would have been had the condensate not collapsed.
We must also check that the radiation from the decays of these squarks does not dissociate the B-balls. 

	       We first consider inflaton decays and 
ignore the radiation energy coming from squark decays. 
Inflaton decays with a constant decay rate $\Gamma$ will produce a 
background radiation energy density with a temperature given by \cite{eu,drt}
\be{t14} T_{r} \approx k_{r} (M_{Pl} H T_{R}^{2})^{1/4} \;\; ; \; \; 
k_{r} = \left(\frac{9}{5 \pi^{3} g(T)} \right)^{1/8}    ~,\ee
where $k_{r} \approx 0.4$ for $g(T) \approx 100$. 
At the time when the B-balls form, corresponding to $H = H_{i}$, where
\be{t14a}  H_{i} \approx 5 \times 10^{-2} 
\left(\frac{|K|}{0.01}\right)
\left(\frac{m}{100\GeV}\right)
\left(\frac{40}{\alpha}\right) \GeV   ~,\ee
the temperature of the background radiation is 
\be{t15} T_{r} \approx    3.4\times 10^{5} 
\left(\frac{|K|}{0.01}\right)^{1/4}
\left(\frac{40}{\alpha}\right)^{1/4}
\left(\frac{m}{100\GeV}\right)^{1/4}
\left(\frac{T_{R}}{10^{3}\GeV}\right)^{1/2}\GeV
~.\ee
The upper bound on $T$ to avoid dissociation, \eq{t5}, may be written as, using \eq{g6},
\be{t15a} T \lae 2\times 10^{5} 
\left(\frac{|K|}{0.01}\right)^{3/8}
\left(\frac{m}{100\GeV}\right)^{3/4}
\left(\frac{10^{3}\GeV}{T_{R}}\right)^{1/4}
\left(\frac{40}{\alpha}\right)^{1/2}
\GeV    ~.\ee
Thus the temperature of the background radiation is essentially 
low enough to evade the dissociation of the B-balls if $T_{R} \lae 10^{3}\GeV$.
 In fact, the 
value of $H$ at which the B-balls can be dissociated will be much less than $H_{i}$. This 
is because at $ H \lae H_{i}$ the squark field expectation value, $\langle |\phi| \rangle$, due to the non-relativistic squarks from the
AD condensate will be large enough to give an effective mass $g\langle |\phi| \rangle$, much larger than $T_{r}$, to any particles
 coupling directly to the B-ball squarks, so Boltzmann suppressing them
 and preventing them from dissociating the B-balls. The value of $\langle |\phi| \rangle$ 
as a function of $H$ is given by $\langle |\phi| \rangle = (H/H_{o}) \phi_{o}$, with $H$ following from \eq{t14}
 and $H_{o} \approx m$. 
$g \langle |\phi| \rangle \gae T$ is satisfied down to 
\be{t16} H \approx 9.6 \times 10^{-10} 
\left(\frac{m}{100 \GeV}\right)^{4/3}
\left(\frac{4.1 \times 10^{14}\GeV}{g \phi_{o}}\right)^{4/3}
\left(\frac{T_{R}}{10^{3}\GeV}\right)^{2/3}\GeV
    ~,\ee
which is much smaller than $H_{i}$ for both
 the d=4 and d=6 condensates. (This assumes that the 
squark condensate has not decayed or thermalized; we will discuss this point below).
The corresponding temperature is given by
\be{16a} T \approx 4 \times 10^{3}  
\left(\frac{m}{100 \GeV}\right)^{1/3}
\left(\frac{4.1 \times 10^{14}\GeV}{g \phi_{o}}\right)^{1/3}
\left(\frac{T_{R}}{10^{3}\GeV}\right)^{2/3}\GeV
~.\ee
Thus dissociation will not occur until $H \ll H_{i}$. 
As a result, dissociation of the B-balls during the 
inflaton dominated period by radiation from inflaton decays will impose no further constraint if the 
reheating temperature satisfies the upper bound coming from dissolution, 
$T_{R} \lae 10^{3-5}\GeV$. 

	      In addition, we should check that the radiation from the
 decay of the low momentum squarks presents no problems. 
To show this, we neglect the radiation coming from the inflaton decays.
The radiation from the squark decays will have a density given by \eq{t13}, where, for 
$g \langle |\phi| \rangle \gae T$, the squark decay rate is approximately given by,
\be{t17} \Gamma \approx \frac{\alpha_{g} m^{3}}{\langle |\phi| \rangle^{2}}       ~,\ee
where $\alpha_{g} = g^{2}/4\pi$ and where $g$ is typically the strong gauge coupling.
This decay rate is time dependent. As a result, the radiation temperature $increases$ as 
$H$ decreases, $T_{r} \propto \rho_{r}^{1/4} \propto H^{-1/4}$. The maximum 
temperature of the radiation from the squark decays will therefore correspond to the 
smallest possible value of $H$, at 
which the energy density in radiation equals that remaining in the low momentum squarks, 
\be{t18} H_{min} \approx \left(\frac{2 \alpha_{g}}{5}\right)^{1/3}
 \frac{m^{5/3}}{\phi_{o}^{2/3}}     ~.\ee
This gives for the maximum temperature of the radiation from squark decays
\be{t19} T_{max} \approx 5.3 \times 10^{3} 
\left(\frac{2 \alpha_{g}}{5}\right)^{1/6}
\left(\frac{m}{100\GeV}\right)^{5/6}
\left(\frac{\phi_{o}}{4.1 \times 10^{14}\GeV}\right)^{1/6} \GeV
~.\ee
This is generally much less than the upper bound on $T$ coming
 from avoiding dissociation, \eq{t15a}, for acceptable values of $T_{R}$.

	     Finally, let us estimate the time at which the squark condensate is thermalized. 
So long as this occurs well after B-ball formation, the B-ball should be protected from 
thermalization by dissociation. Assuming that $g \langle |\phi| \rangle \gae T$, the largest thermalization 
rate possible will be of the order of 
\be{t20} \Gamma \approx \frac{\alpha_{g} T^{3}}{\langle |\phi| \rangle^{2}}       ~.\ee
Thermalization will occur once $\Gamma \gae H$, which, assuming inflaton domination, occurs 
once $H$ is smaller than $H_{therm}$, where
\be{t21} H_{therm} \approx 4 \times 10^{-4} \alpha_{g}^{4/9} 
\left(\frac{T_{R}}{10^{3}\GeV}\right)^{2/3}
\left(\frac{4.1 \times 10^{14}\GeV}{\phi_{o}}\right)^{8/9}
\left(\frac{m}{100 GeV}\right)^{8/9}\GeV
   ~.\ee
For the d=6 condensate this is much smaller than the value at which the B-balls form, $H_{i}$,
 when $T_{R} \lae 10^{3-5}\GeV$ and $|K|$ is in the range 0.01 to 0.1. For the
d=4 condensate, on the other hand, for which typically
 $\phi_{o} \approx 10^{10}\GeV$, it can be larger than $H_{i}$,
 depending on the values of $T_{R}$ and $|K|$. Thus it is possible that the d=4 squark condensate
 could be thermalized before the B-balls form. Let us finally note that the
 squark condensate will thermalize before the reheating temperature is reached (corresponding to 
$H(T_{R}) \approx 2 \times 10^{-12} (T_{R}/10^{3}\GeV)^{2}\GeV$) 
for all values of $T_{R}$ consistent with B-balls surviving thermalization.  

    To conclude, we find that B-balls can survive thermalization both during and after the inflaton dominated 
era so long as the reheating temperature is sufficiently small, $T_{R} \lae 10^{3-5}\GeV$ for $|K|$ in the 
range 0.01 to 0.1. The strongest constraint comes from dissolution of the B-ball by thermalization and diffusion
of charge from the soft edge.

\section{B-ball decay} 

	   Assuming that the reheating temperature after inflation is sufficiently low that the 
B-balls survive thermalization, the next question to be considered is that of their decay temperature. 
Typically the LSP freeze-out temperature is $T_{fr} \approx m_{\chi}/20$. Therefore if the LSP is
not very heavy, say $m_{\chi} \lae 200\GeV$, then we must require that the B-balls decay at
temperatures less than around 10 GeV, in order that they decay after
the freeze-out of the LSPs, with the LSPs from B-ball decay dominating
the relic density. We must also ensure that there is no subsequent annihilation of
 the LSPs coming from B-ball decays, in order to maintain the direct relationship
 between the baryon and dark matter particle number densities.   

\subsection{Decay temperature}

In the previous discussions of B-ball decay, the decay rate 
has been estimated from 
the model of reference \cite{cole2}, which assumes that there is a single scalar 
coupled to a light fermion. 
In this case the decay to light fermions is proportional to the area of the 
B-ball (decay within the 
volume of the B-ball being blocked by the Pauli principle once the Dirac 
sea is filled),
 with an upper bound on the decay rate being given by \cite{cole2}
\be{d1} \left|\frac{dB}{dt}\right|_{fermion} \leq  \frac{\omega^{3} A}{192 
\pi^{2}}   ~,\ee
where $A$ is the area of the B-ball. The upper bound is likely to be 
saturated for 
B-balls with $\phi(0)$ much larger than m. However, in realistic SUSY
 models, there will
be several scalar fields and in general we would not expect 
that the condensate scalar will correspond to the 
lightest scalar. Since the decay to scalar fields within the volume is 
not blocked by the 
Pauli principle, it is possible that the decay to light scalars could be significantly 
enhanced relative to the decay to fermions. 
However,  the decay to light scalars will only be possible near the edge of 
the thick-walled B-ball. This is because particles coupling directly to the condensate scalars will
gain a large effective mass from $\langle \phi \rangle$ inside the B-ball. As a result, decay to light scalars will occur only 
via loop diagrams with rates suppressed by this large effective mass. 

    In the previous subsection we derived a bound on the reheating temperature
from B-ball stability. Let us now derive a general upper bound on the 
reheating temperature from the requirement that 
B-balls decay below the LSP freeze-out temperature. 
Let $f_{s}$ be the possible enhancement factor of the scalar
decay rate over the fermion decay rate, 
\be{d2} \frac{dB}{dt} =
f_{s} \left(\frac{dB}{dt}\right)_{fermion}
~.\ee 
The B-balls will decay once the decay rate is larger than $H$. This gives for the decay
temperature $T_{d}$
\be{d3} T_{d} \approx 
\left( \frac{f_{s}\omega^{3}R^{2}M_{Pl}}{48 \pi k_{T} B}\right)^{1/2} 
 \approx 0.06  \left(\frac{f_{s}}{|K|}\right)^{1/2} 
\left( \frac{m}{100\GeV} \right)^{1/2}
\left( \frac{10^{20}}{B} \right)^{1/2} \GeV
~.\ee
Using \eq{g6}, we find that 
requiring that $T_{d}$ is less than 10 GeV imposes an upper limit on 
the reheating temperature
\be{d4} T_{R} \lae 5 \times 10^{8} \;
\frac{|K|^{3/2}}{f_{s}}
\left( \frac{100\GeV}{m} \right)^{2}
\left( \frac{40}{\alpha} \right)^{2}
\left( \frac{T_{d}}{10\GeV}
\right)^{2}\left(\frac{\eta_B}{10^{-10}}\right)
 \GeV
~.\ee
Thus, with $|K| \approx (0.01-0.1)$, 
we need $T_{R} \lae (5 \times 10^{5}-2 \times 10^{7})f_{s}^{-1}\; \GeV$ 
for the B-balls to decay at a temperature below 10\GeV. 

	  We next consider the possible enhancement of the decay rate to light
scalars over the decay rate to light fermions. We first note that, for the 
particular case of the d=6 $\udd$ direction, it is quite possible that there will be no scalar decay mode enhancement.
The condensate scalar in this case is a linear combination
of right-handed squarks. For models with universal soft SUSY breaking terms at large renormalization mass scales, 
the solution of the renormalization group equations implies that 
the left-handed squark masses will typically be heavy compared with the right-handed squark masses
at low mass scales \cite{nilles}. The Higgs scalar masses can also be heavy compared with the right-handed squark 
masses, depending on the $\mu$ parameter. On the other hand, 
in models with universal soft SUSY breaking terms at the large mass scale,
the sleptons will typically
be lighter than the right-handed squarks.
However, even if slepton masses or Higgs masses 
were less than
the right-handed squark masses,  
any condensate squark decay to sleptons or Higgs bosons 
would also have to involve a quark in the final state, together with a gaugino-Higgsino or lepton.
So this would effectively involve pair producing light fermions in the final state, which would be suppressed by 
the Pauli principle as for the two fermion decay process.  
 The only exception to this 
would be if the mostly right-handed stop was sufficiently light 
compared with the other right-handed squarks that decay 
of the condensate squark to a stop plus a 
Higgs boson became kinematically possible. This is a model-dependent possibility. 
Therefore it is quite possible that decay of the linear combination of right-handed squarks, which
constitutes the condensate scalar in the d=6 $\udd$ direction, to pairs of lighter scalars
 will be kinematically forbidden and that $f_{s}$ will be effectively equal to 1. 
In this case it is sufficient that $T_{R}$       
be less than about $10^{5-7}\GeV$ for the B-balls to decay below 10 GeV.

	     More generally (for example. for the case of a B-ball made of left-handed squarks or 
sleptons, such as for the d=6 directions based on $d^{c}QL$ or $e^{c}LL$), 
we would expect a decay mode to pairs of light scalars to exist. 
In this case we can estimate the largest 
possible enhancement factor using the Gaussian thick-wall Ansatz. 
Within the B-ball, for values of $\phi$ much larger than $m$, 
the lowest possible dimension operator which 
could allow the condensate scalars to decay at one-loop to light particles is the d=5 operator
\be{d5} \frac{1}{M} \int d^{4}\theta \phi \chi^{\dagger} \eta           ~,\ee
where $\chi$ and $\eta$ represent the light particles and $M \approx g\phi$, where $g$ is the 
coupling of the heavy particles to $\phi$. 
The rate of decay of the condensate scalars to light scalars will then be 
\be{d6} \frac{dB}{dt} = -\int \omega \phi^{2}(r) \Gamma(r) 4 \pi r^{2} dr ~,\ee
where $\omega \phi^{2}(r)$ is the charge density within the B-ball and where
\bea{d7} \Gamma(r) &\approx& \frac{\alpha^{2} m^{3}}{\phi^{2}} 
\; ; \;\;\; g\phi > m ~\nn
&\approx& \alpha m \; ; \;\;\; g\phi < m    ~,\eea
with $\alpha = g^{2}/(4 \pi)$. (For simplicity we consider a single coupling constant $g$). 
Let $r_{*}$ be the radius at which 
$ \phi(r) = m/g$. Then the
 largest contribution to the decay rate will come from a region of width 
$\delta r \approx R^{2}/(4 r_{*})$ around $r_{*}$, over which $\phi$ has a roughly constant 
value $ \phi \approx m/g$, where 
\be{d9} r^{*} = \gamma R  \;\; ; \;\;\;\;\; \gamma = ln^{1/2}\left( \frac{g \phi(0)}{m} 
\right)     ~.\ee
From \eq{d6} this gives a rate,  
\be{d10} \frac{dB}{dt} \approx 
- 4 \pi \alpha \omega m \left(m^{2}\alpha \int_{0}^{r_{*}} dr r^{2} 
+ \phi^{2}(0) \int_{r_{*}}^{\infty} dr r^{2} e^{-\frac{2 r^{2}}{R^{2}}} \right) ~\ee
\be{d10a} \approx -\frac{12 \pi \gamma}{|K|^{1/2}} 
\left(1 + \frac{\gamma^{2} g^{4}}{3 \pi}\right)
\left(\frac{dB}{dt}\right)_{fermion}
~.\ee 
where we have used $\omega \approx m$ (see Appendix).
  For the thick-walled B-ball, for typical values of the parameters, $g\phi(0)/m 
\approx (0.1-0.01)B^{1/2}$ and so 
$\gamma \approx 4.5$. Thus we find that the enhancement factor is typically given by
\be{d11} f_{s} \approx \frac{170}{|K|^{1/2}}
\left(1 + 2.1 g^{4}\right)      ~.\ee
Thus, if $g$ is less than 1, then for $|K| \approx 0.01-0.1$ we expect an enhancement factor
not much larger than about $10^{3}$. (For $g$ less than 1 most of the 
enhancement factor comes from unsuppressed tree level decays occuring at $r > r_{*}$). 
This would impose an upper bound on the reheating 
temperature of about $10^{3-5}\GeV$ in order to have B-ball decay at a temperature less than $10\GeV$. 
Therefore, if it is possible for the condensate scalars to decay to light scalars, then the upper bound on 
$T_{R}$ will be much tighter, although it will not be very
 much different from that coming from the requirement 
that the B-balls survive thermalization by dissolution.

    Thus, if $T_{R} \lae 10^{3}\; (10^{5})\GeV$ for $|K| = 0.01\; (0.1)$ , then we expect
 that the B-balls will typically survive thermalization $and$ will decay at a temperature below 10
GeV. In this case we may be able to understand the baryon to dark matter 
ratio as being 
a consequence of B-ball decays. Just how much dark matter is produced 
will depend on the details of the thermalization and annihilation of the LSPs coming 
from the decay of the B-balls. 
It should be emphasized that the requirement that B balls evade 
thermalization and that
they decay at a low enough temperature to account for the baryon to 
dark matter ratio both
point to a relatively low reheating temperature after inflation. We will see 
that this is also preferred by the 
naturalness of the observed baryon asymmetry in this scenario.

\subsection{LSP dark matter from B-ball decay}

A typical mass of a B-ball would be $10^{20}(10^{3}GeV/T_{R})\GeV $. When they decay at
$t\approx \Gamma_B^{-1}$, the average distance between two B-balls is
$l_{B} \approx 5 \times 10^{-8} (T_{d}/10GeV) l_{H}$,
where $l_H$ is the
horizon distance and $T_{d}$ is the decay temperature 
(and assuming that the baryon number of the Universe is mainly due to
B-ball decay). As we discussed in the Introduction, there will be
$N_\chi\gae 3$ LSPs produced per baryon number, so that as a reference
number we take the total number of LSPs produced in each B-ball decay 
to be $N_{\rm LSP}^{\rm tot}\approx 10^{20}(10^{3}GeV/T_{R})$. 
Because the B-ball decay 
gives rise to an overabundance of LSPs with respect to the relic
density, we must check that the produced LSPs will not get annihilated, 
both during the initial decay of the B-balls and subsequently once the 
LSP density is smoothed out. 

       Let us consider the decay of a single B-ball. The produced LSPs will 
collide with the weakly interacting particles in the background and
after a few  collisions \cite{sirkka} will locally settle into a kinetic
equilibrium (some radial bulk motion might still exist). Thus they soon become
non-relativistic as $T_d$ is by assumption less than the freeze-out
temperature of the LSPs, which we may take to be about $m_\chi/20$ \cite{susydm}.
The rough
freeze-out condition for LSPs initially in thermal equilibrium is
\be{lspfreeze}
n_{\rm LSP}\langle \sigma_{\rm ann}v\rangle \approx H_f {m_\chi\over T_f}~,
\ee
where $\sigma_{\rm ann}$ is the LSP annihilation cross-section and 
the subscript $f$ refers to the freeze-out values. LSPs annihilate to
light fermions. The thermally averaged cross section can be written as $\langle
\sigma_{\rm ann}v\rangle =a+bT/m_\chi$, where $a$ and $b$ depend on the couplings and
the masses of the light fermions \cite{susydm}. 
In the following we will consider the case of a light neutralino, $m_{\chi} < m_{W}$, 
and we will neglect the final state fermion masses. (This would be appropriate for the case of 
pure BBB ($f_{B} =1$), for which $m_{\chi} \lae 67GeV$). In this case $a=0$ and
$\langle \sigma_{\rm ann}v\rangle\propto T$. In that case one obtains from \eq{lspfreeze}
that $b \approx H m_{\chi}^{2} T_{f}^{-2} n_{f}^{-1}$, where 
\be{eqdist}
n_f=\frac{1}{(2\pi)^{3/2}}(m_\chi T_f)^{3/2}e^{-m_\chi/T_f}\approx 
1.46\times 10^{-12}
m_\chi^3~,
\ee
is the freeze-out LSP density and where we have used $T_f\approx m_\chi/20$.

	     The LSPs produced in the decay of the B-ball will spread out by a random
walk with a rate $\nu$ determined by the collision frequency divided by the
thermal velocity $v_{th} \approx \sqrt{T/m_\chi}$,
which reads
\be{collfreq}
\nu^{-1}\approx (\langle \sigma_{sc} v_{\rm rel}\rangle n_{rad}/v_{th}^{2})\approx 
    g(T) G_{F}^{2} m_{\chi} T^{4}   \;,
\ee
where $\sigma_{sc} \approx 36 G_{F}^{2} T^{2}/\pi$ is the scattering cross-section with thermal particles, 
$n_{rad}$ is the density of particles in radiation and
 we have taken $v_{\rm rel}\approx 1$.
It is very likely that the decay is spherically symmetric. Thus the 
number of
LSPs created at $t=0$ subsequently form a Gaussian distribution around
the decaying B-ball with
\be{lsp1}
\frac{dN_{\rm LSP}(r,t=0)}{dr}=\sqrt{\frac{2}{\nu t\pi}}e^{-r^2/(2\nu t)}\;.
\ee
Folding in the rate by which the LSPs are produced,
\be{prodlsp}
{dN_{\rm LSP}(t)\over dt}=\Gamma_BN_{\rm LSP}^{\rm tot}e^{-\Gamma_Bt}\;,
\ee
and integrating over time we find that the radial distrubution of the LSPs at
time $t$ is given by
\be{ntot}
\frac{dN_{\rm LSP}^{\rm tot}(r,t)}{dr}=\int_0^tdt'
\sqrt{\frac{2}{\nu (t-t')\pi}}
\Gamma_BN_{\rm LSP}^{\rm tot}e^{-\Gamma_Bt'}
e^{-r^2/(2\nu (t-t'))}\;.
\ee
The largest contribution to the integral comes from the region
$t'\approx \Gamma_B^{-1}\gg \nu$ so that in the leading approximation
\eq{ntot} reads
\be{lspappro}
\frac{dN_{\rm LSP}^{\rm tot}(r,t)}{dr}
\approx N_{\rm LSP}^{\rm tot} \left({2\Gamma_B\over \pi\nu x}\right)^{1/2}
e^{-r^2\Gamma_B/(2\nu x)}\;,
\ee
where we have written $t=\Gamma_B^{-1}x$ with $x\sim {\cal O}(1)$. Thus there is
a "central region" 
of radius 
\be{rcent} \overline{r} \approx \left( \frac{\nu x}{\Gamma_{B}}\right)^{1/2}
~\ee
within which the mean LSP number density is given by
\be{ndens}\overline{n}_{LSP}(\overline{r}) \approx \frac{3 N_{LSP}^{tot}}{4 \pi \overline{r}^{2}}
\left(\frac{2 \Gamma_{B}}{\pi \nu x}\right)^{1/2}
~,\ee
and outside of which it is exponentially suppressed.

     In the central region annihilation is significant if
$\bar n_{\rm LSP}\langle \sigma_{\rm ann}v_{\rm rel}\rangle \gae H$.
Neglecting the possible change
in the  light degrees of freedom below $T_f$ so that $H/H_f=T^2/T_f^2$, one finds 
that significant annihilation means simply that $\bar n_{\rm LSP}(r)\gae (T/m_{\chi}) n_f $.
From \eq{ndens} this happens when (we now take $x\approx 1$)
\be{rcond}
\overline{r} \lae r_c\equiv 1.3\times 10^9 {T m_\chi^{-5/4}\over 
\GeV^{3/4}}
\left(\frac{g(T)}{100}\right)^{1/4}
\left({N_{\rm
LSP}^{tot}\over 10^{20}}
\right)^{1/2}~.
\ee
A large fraction of the LSPs produced in the B-ball decay will thus be destroyed
if the critical distance $r_c$ is greater than the size of the central region,
$ \overline{r} \equiv \sqrt{\langle r^2\rangle}=(\nu/\Gamma_B)^{1/2}$. Otherwise the build-up of LSP
density is slow so that the main bulk of the LSPs will have time enough to escape
from the vicinity of the B-ball. We find that
\be{rrrat}
{r_{c}\over \overline{r}} = 1.6\times 10^{-4}\left({T_{d}^{4}\over
m_{\chi}^{3/4} \GeV^{13/4}}\right)\left({N_{\rm LSP}^{tot}\over 10^{20}}\right)^{1/2}
\left(\frac{g(T)}{100}\right)^{3/4}
~.\ee
Thus annihilation is insignificant provided
\be{noann}
T_d\ll 21 \left({m_\chi \over 100 \GeV}\right)^{3/16}\left( {10^{20}\over N_{\rm
LSP}^{tot}}\right)^{1/8}
\left(\frac{100}{g(T)}\right)^{3/16}
~\GeV~.
\ee
Because it is likely that there remains some radial
bulk motion even after LSP equilibration, the actual limit on $T_d$ is bound to be
somewhat less stringent. In any case, we may conclude 
that typically most of the LSPs will survive if the B-ball decay
temperature is less than a few GeVs. 

	     The above discussion applies to the case of a single B-ball. However, we must also
ensure that there is no subsequent LSP annihilation once the LSP density is smoothed out. 
This can be significant because there is typically more than one B-ball within the central region
of a given B-ball, so it is not sufficient to check that the LSP's can escape from a single B-ball
without significantly annihilating. Since the essence of the natural explaination of the baryon to 
dark matter ratio via B-ball decay is the similarity of the number of LSPs and baryons 
produced by B-ball decay, we must ensure that the number density of LSP's is not 
significantly reduced by annihilations, as well as ensuring that the thermal relic density 
of LSP's is small compared with that from B-ball decay. 

		      The upper limit on the number of LSP's at a given temperature is 
\be{upperlimit} n_{limit}(T) \approx \frac{H}{\langle \sigma v \rangle_{ann}}   ~.\ee
For an annihilation cross-section dominated by $b$ at $T_{d}$, 
$${ n_{limit}(T) \approx (g(T_{f})/g(T))^{1/2}(T_{f}/T)^{2} n_{relic}(T)},$$
where $n_{relic}(T)$ is the thermal relic density at $T$ 
\cite{susydm}, 
\be{nrelic} n_{relic}(T) \approx  
\left(\frac{g(T)}{g(T_{f})}\right)
\left(\frac{T}{T_{f}}\right)^{3} 
\left( \frac{H}{\langle \sigma v \rangle_{ann}}\right)_{T_{f}}   ~.\ee
Thus in order to have a natural explaination of the baryon to dark matter ratio we must have
\be{condition} n_{relic}(T_{d}) < n_{LSP}(T_{d}) < 
\left(\frac{g(T)}{g(T_{f})}\right)^{1/2}
\left(\frac{T_{f}}{T_{d}}\right)^{2} n_{relic}(T_{d})     ~,\ee
where
\be{nlsp} n_{LSP}(T_{d}) \approx N_{\chi} n_{B} = 1.3 \times 10^{-5} 
\left(\frac{T_{d}}{10\GeV}\right)^{3}
\left(\frac{g(T_{d})}{100}\right)
\left(\frac{N_{\chi}}{3}\right)
\left(\frac{\eta_{B}}{10^{-10}}\right)
\GeV^{-3}
~\ee
 is the LSP density from B-ball decay (assuming that the B asymmetry comes mostly from B-ball decay).
For a reasonable range of LSP densities to exist, the B-balls must decay sufficiently well 
below the LSP freeze-out temperature. 

	  For the $b$ dominated annihilation cross-section, with $m_{\chi}/T_{f} \approx 20$, 
we find that the condition $n_{LSP}(T_{d}) > n_{relic}(T_{d})$ is $T_{d}$ independent
and generally satisfied, with
 \be{relicratio} n_{LSP}(T_{d}) \approx 23 
\left(\frac{g(T_{f})}{100}\right)
\left(\frac{N_{\chi}}{3}\right)
\left(\frac{\eta_{B}}{10^{-10}}\right)
n_{relic}(T_{d})
~.\ee 
The condition that $n_{LSP}(T_{d}) < n_{limit}(T_{d})$ is satisfied if 
\be{limit} T_{d} \lae 1.2
\left(\frac{3}{N_{\chi}}\right)^{1/2}
\left(\frac{100}{g(T_{f})}\right)^{1/2}
\left(\frac{10^{-10}}{\eta_{B}}\right)^{1/2}
\left(\frac{m_{\chi}}{100\GeV}\right)
\GeV
~.\ee
Thus we can conclude, for the case of a $b$ dominated annihilation cross-section, 
appropariate for a light neutralino, that the baryon to dark matter ratio can be naturally
explained by B-ball decay if the B-balls decay at a temperature less than about $1\GeV$. 
This in turn requires that $T_{R} \lae (5 \times 10^{3}- 2 \times 10^{5})f_{s}^{-1} \GeV$ for
$|K| \approx (0.01-0.1)$. We note that this may favour the $\udd$ direction, for which 
$f_{s}$ can be equal to 1, over directions along which the B-balls can decay to light scalars, 
for which $f_{s} \approx 10^{3}$, so imposing a much 
tighter upper bound on the reheating temperature. 

\section{Application to the $\udd$ and $\uude$ directions}
In this section we apply the results of the 
previous sections to the $\udd$ and 
$\uude$ squark directions. We first note that the upper bounds on the
 reheating temperature following from the 
requirement that the B-balls are not thermalized by
 dissolution and that they decay at a sufficiently 
low temperature are only very weakly dependent on
 the details of the scalar potential i.e. logarithmically via 
$\alpha(\lambda_{o})$. Thus these bounds will be effectively independent of d. 

       The initial value of the field when the condensate oscillations begin, for the d=6 case, 
is given by, 
\be{p1} \phi_{o} = 5.8 \times 10^{14} \lambda^{-1/4} 
\left(\frac{m}{100\GeV}\right)^{1/4} \GeV     ~\ee
and for the d=4 case by
\be{p2} \phi_{o} =  3.2\times 10^{10}
\lambda^{-1/2} 
\left(\frac{m}{100\GeV}\right)^{1/2} \GeV     ~.\ee
Thus the lower limit on the reheating temperature, following from the requirement of 
maximum efficiency for the d=6 case, is given by
\be{p3} T_{R} \gae 0.23 \; \lambda^{1/2} 
\left(\frac{m}{100\GeV}\right)^{1/2} \GeV   ~,\ee
where the natural value of $\lambda$ for d=6, assuming that the strength of the
 non-renormalizible interactions is set by $M_{p}$, is around 0.003.               
This is easily satisfied for reheating temperatures less than the upper bounds     
from B-ball thermalization and decay. For the d=4 case, the upper bound for maximum 
efficiency is given by
\be{p4} T_{R} \gae 8 \times 10^{7} \lambda \;\GeV   ~,\ee
where $\lambda$ is naturally around 0.1 for d=4.
This cannot be satisfied if the reheating temperature is low enough for the B-balls to survive
thermalization, $T_{R} \lae 10^{3-5}\GeV$.

     Thus efficient B-ball formation is quite natural for the d=6 $\udd$
 direction. As noted in the previous section, this direction is particularly favoured 
as $f_{s} = 1$ is possible, allowing a relatively large reheating temperature to be compatible
with similar baryon and dark matter particle number densities.
Therefore, so long
as the reheating temperature after inflation is
no greater than around $10^{3-5}\GeV$, the baryon
 to dark matter ratio can be naturally accounted for via B-ball decays 
in this case. 

	      For the d=4 direction, the B-balls do not form efficiently 
 However, for the d=4 $\uude$ direction B-L is conserved. Therefore, in this case only the baryon number trapped in the B balls will survive. Thus 
it may appear that the observed B asymmetry can be obtained by simply having a much larger B asymmetry 
in the initial condensate than the present B asymmetry. However, for the d=4 condensate, no 
B-balls will survive thermalization if we can generate the observed baryon asymmetry.
The maximum possible 
asymmetry will simply correspond to the number density of $\phi$ scalars in the condensate. When the 
condensate scalar starts oscillating at $H \approx m$, this is given by $n_{B\;o} \approx m \phi_{o}^{2}/2$. The baryon to entropy
ratio is related to $n_{B\;o}$ by 
\be{p5} \eta_{B} \approx 6 \times 10^{-33} \left(\frac{T_{R}}{10^{9}\GeV}\right) 
\left(\frac{100 \GeV}{m}\right)^{2} 
n_{B\;o}        ~.\ee
Thus, with $\phi_{o}$ for d=4 given by \eq{p2}, we find that the maximal baryon to entropy ratio initially is 
$\eta_{B} \approx 6 \times 10^{-10} \lambda^{-1} (T_{R}/10^{9}\GeV) \lae 10^{-8}$. 
Therefore 
the d=4 condensate can only account for the observed B asymmetry if the asymmetry is not 
very much below the maximum possible for a d=4 condensate,
 which requires a high reheating temperature, $T_{R} \gae 10^{7}\GeV$.
In this case no B-balls will survive thermalization. Thus, in general, the d=4 condensate 
will not be able account for the baryon to dark matter ratio via B-ball decay. 

	    For the case of the d=6 condensate it is important to note that, as well
 as allowing for the baryon to dark 
matter ratio to be explained via B-ball decay, 
 a low reheating temperature is 
also the most natural from the point of view of the observed value of the B asymmetry. 
The observed B asymmetry for the d=6 case,
 assuming that B is conserved from the time of condensate formation until the present, is 
given by 
\be{p6} \eta_{B} \approx 0.2 \left(\frac{T_{R}}{10^{9}\GeV}\right)
\left(\frac{100 \GeV}{m}\right)^{1/2}
\left(\frac{0.003}{\lambda}\right)^{1/2} \delta_{CP}    ~,\ee
where $\delta_{CP}$ is the CP violating phase (the phase of the A-term) responsible for the B asymmetry in the initial 
condensate. 
Thus we see that if $T_{R}$ is small compared with $10^{9}\GeV$ then we do 
not need to have a CP violating phase that appears unnaturally small compared with 1 in order to account for the 
observed baryon asymmetry, $\eta_{B} = (3-8) \times 10^{-11}$ \cite{sarkar}.
 For example, $T_{R} \lae 10^{3}\GeV$ would allow the 
present B asymmetry to be explained by a phase greater than or of the order of $10^{-3}$.
We also note that this is most 
natural if the CP violation responsible for the B asymmetry enters 
through the A-terms  in the scalar potential, rather than being set by
an initial random phase in the AD field due to de Sitter fluctuations 
during inflation, which would tend to give $\delta_{CP} \approx 1$.
This in turn requires that order $H$ corrections to the A-terms exist,
in order to damp out any large random phase, which may serve as an
 important constraint on inflation models.
 Thus both the baryon to dark matter ratio and the observed
B asymmetry can be most naturally explained by having the B asymmetry originate
 from a squark condensate along a d=6 D-flat direction of the scalar 
potential with a low reheating temperature after inflation. The $\udd$ direction is 
particularly favoured, typically allowing a larger reheating temperature than 
the other d=6 directions in the MSSM by evading the 
large scalar mode enhancement of the B-ball decay rate.

	   This all assumes that the B asymmetry from the d=6 
condensate is conserved. In fact, this is likely to be effectively
true in general, even if there were additional L-violating interactions, 
since the B-balls in this case are likely to form with
a high efficiency, protecting most of the asymmetry.
 Only if the reheating temperature were very low,
less than around 1 GeV, would there be a reduction in the efficiency of
B-ball formation along the d=6 direction. However, from \eq{p6},  the
reheating temperature cannot be much less than 1 GeV and still be able
to account for the  observed asymmetry, even if all the asymmetry could
survive. 
 
\section{Conclusions}

	    We have considered the possibility of accounting for the 
baryon to dark matter ratio via B-ball decay in the MSSM. 
 We have considered the constraints on the
reheating temperature after inflation, 
in the context of a simple cosmological scenario,
following from the requirements that the B-balls can survive
thermalization, decay sufficiently long after LSP freeze-out and
naturally account for the observed B asymmetry. 

	 For the case of a d=4 AD
condensate, we find that, in general, the reheating temperature after
inflation must be too high for the B-balls to have survived
thermalization. In contrast, for the case of the d=6 $\udd$ condensate, 
a perfectly consistent cosmological scenario emerges.
The requirements that the B-balls survive thermalization, can
 form efficiently and so naturally account for the 
baryon to dark matter ratio and can decay sufficiently below
 the LSP freeze out temperature without subsequent annihilation 
are all satisfied for 
typical values of the radiative correction to the scalar potential
if the reheating temperature is between 1$\GeV$ and $10^{3-5}\GeV$.
 A low reheating temperature is also preferred if 
the B asymmetry is to originate from a d=6 AD condensate without
 requiring an unnaturally small CP violating phase. The $\udd$ direction 
is particularly favoured, allowing a larger reheating temperature than the 
other d=6 directions in the MSSM by avoiding the scalar mode enhancement of the B-ball decay rate. For the case of pure B-ball Baryogenesis, 
where all of the baryon asymmetry comes from B-ball decay, the LSP cannot have a mass
greater than 67GeV. 

	     It seems remarkable that with only the fields of MSSM, together with plausible 
non-renormalizible corrections and period of primordial inflation, we are able to
 naturally account for a baryon to dark matter ratio of the observed form. 
Given that following inflation there will typically be an ensemble of domains, much larger than the observed Universe, in which the MSSM scalars take on all possible initial values along F- and D-flat directions, it seems certain that there will be at least some domains of the Universe in which the scalar fields will have initial values along the $\udd$ direction. All we then require, in order to account for the baryon to dark matter ratio, is that the reheating temperature be sufficiently low. In a sense, this may be considered a prediction of this scenario, providing a strong motivation for inflation models with low reheating temperatures. The requirement of a low reheating temperature will impose a significant constraint on inflation models, tending to favour inflaton  
candidates which are light and/or highly decoupled from the MSSM or other light fields. 

	      There are several issues which remain to be discussed. The details of the 
dark matter density coming from B-ball decay and the resulting constraints on the 
MSSM parameter space should be investigated for general neutralino candidates.
 The low B-ball decay temperature may also have interesting cosmological
 implications for the QCD phase transition and nucleosynthesis.
Ultimately, we would hope to be able to provide a completely consistent model of B-ball Baryogenesis and the baryon to dark matter ratio in the context of a realistic SUSY inflation model which can satisfy the low reheating temperature constraint.
 Such a model would provide a basis for
 a complete cosmological scenario which
 would be able to naturally account for all cosmological
 observations (baryon number, dark matter and primordial fluctuations)
 and which would require no low energy physics beyond that of the MSSM. 

\subsection*{Acknowledgements}   This work has been supported by the
 Academy of Finland and by a Marie Curie Fellowship under EU contract number 
ERBFM-BICT960567.

\section*{Appendix. Some aspects of B-ball solutions}

	       In this Appendix we discuss some properties of numerical B-ball solutions and 
show that, for the purposes of discussing the thermalization and decay
 of the B-balls, a Gaussian Ansatz is a reasonable approximation to the exact B-ball solution. 

		From the point of view of cosmology
 and phenomenology, the important quantites 
are the energy and radius of the B-ball
as a function of its charge $B$. The
 B-ball solution is of the form 
\bea{q-ball}   \Phi = \frac{\phi(r)}{\sqrt{2}} e^{i \omega t}  ~.\eea
The energy and charge of the B-ball are then given by \cite{cole,k}
\bea{energy}  E = \int d^{3}x \left[ \frac{1}{2}
 \left(\frac{\partial \phi}{\partial r}\right)^{2} 
+ U(\phi)  \right] + \frac{1}{2} \omega B
~\eea
and 
\bea{charge}  B =   \int d^{3}x  \omega \phi^{2}  ~,\eea
where we rescale $B$ such that $B = +1$ for the scalars.
The equation of motion for a B-ball of a fixed value of $\omega$ is given by \cite{cole}
\bea{eqball}  \phi^{''} 
+ \frac{2}{r} \phi^{'} =  \frac{\partial U(\phi)}{\partial \phi}
- \omega^{2} \phi   ~\eea
where $\phi^{'} = d\phi/dr$. 
We require a solution such that $\phi(0) = \phi^{'}(0) = 0$ and $\phi \rightarrow
0$ as $r \rightarrow \infty$. This corresponds to a tunnelling solution for the potential 
$-\overline{U}(\phi)$ \cite{cole}, where 
\bea{ubar} \overline{U}(\phi) = U(\phi) - \frac{\omega^{2}}{2} \phi^{2} ~.\eea 
In practice, when obtaining numerical solutions, we vary $\phi(0)$ with these boundary conditions 
until the correct form of solution is obtained 
for a given $\omega$. The energy and charge of the
solution are then calculated using the above expressions. 

	  For the d=6 direction, the B-ball equation is given by 
\bea{a1}   \phi^{''} + \frac{2}{r} \phi^{'} = -  \omega_{o}^{2} \phi 
+  m^{2} \phi K \log \left( \frac{\phi^{2}}{M^{2}}\right)
 + \left(\frac{10 \lambda^{2}}{32}\right)
\frac{\phi^{9}}{M_{p}^{6}}    ~,\eea
where $\omega_{o}$ is defined by 
\bea{a2} 
\omega_{o}^{2} = \omega^{2} -  m^{2} \left( 1 +
 K \right) 
~\eea
and $M$ is of the order of the scalar field within the B-ball.

	 For the case of thin-walled B-balls, the initial value of $\phi$ is very close 
to $\phi_{c1}$, the value of $\phi$ for which the right-hand side of \eq{a1} 
vanishes.
In this case $\phi$ will remain close to $\phi_{c1}$, up to a radius
of the order of $\omega_{o}^{-1} log\left(\phi_{c1}/\delta \phi(0)\right) $ where 
$\delta \phi(0) = (\phi_{c1}-\phi(0))$. It will then decrease to zero over a distance 
$\delta r \approx \omega_{o}^{-1}$, corresponding to the width of the wall of 
the B-ball. The radius of the thin-walled B-ball can be made arbitrarily large by
 choosing $\delta \phi(0)$ small enough. 

       For the case of thick-walled B-balls, the initial value of $\phi$ can be much smaller than $\phi_{c1}$.
In this case the non-renormalizible terms may be neglected. In general, the right-hand side (RHS) of the 
B-ball equation vanishes for three values 
of $\phi$, which correspond to $\phi_{c1}$, $\phi_{c2}$ and zero. $\phi_{c2}$ corresponds to the point at which, assuming that 
the non-renormalizible terms can be neglected, the first two terms on the RHS cancel,
\bea{a3} \phi_{c2} = M exp\left( \frac{\omega_{o}^{2}}{2 K m^{2}}\right)   ~.\eea
$\phi_{c2}$ is an attractor, in the sense that if $\phi(0)$ is close to $\phi_{c2}$ it will
tend towards $\phi_{c2}$ as $r$ increases. We can now understand qualitatively how
the thick-walled B-ball solution works. Suppose that  initially $\phi(0)$ is large 
compared with $\phi_{c2}$ but small compared with $\phi_{c1}$. Then 
as $r$ begins to increase the RHS of the B-ball equation will be approximately 
$-\omega_{o}^{2} \phi$. Thus for small enough $r$ the solution of the B-ball equation is
\bea{a4} \phi(r) \approx \frac{\phi(0)}{\omega_{o}}
 \frac{Sin( \omega_{o} r)}{r}       ~.\eea 
This solution would become negative for $r \gae \omega_{o}^{-1}$. 
However, as $\phi$ approaches $\phi_{c2}$, the attractive nature of this value of 
$\phi$ alters the solution and, for one particular value of $\phi(0)$ for a given $\omega_{o}$, 
causes the solution to "level out" and smoothly tend to zero as $r \rightarrow \infty$. This 
gives the thick-wall B-ball solution for a given $\omega_{o}$. 

	       Numerically, we find the following properties for the thick-walled B-ball solutions; 
\bea{a5} R_{c} \approx \frac{(1.4-1.6)}{|K|^{1/2} m}    ~,\eea
where $R_{c}$ is defined as the radius within which 90$\%$ of the B-ball energy 
is found, and 
\bea{a6} \omega_{o} \approx (3-4) |K|^{1/2} m               ~.\eea
Since typically $|K|$ is small compared with 1, we see that $\omega \approx m$.

		We next show that a Gaussian Ansatz is a physically reasonable appoximation to the 
thick-wall B-ball solution.  
If we insert the Gaussian Ansatz,
\bea{a7}  \phi(r) = \phi(0) e^{-\frac{r^{2}}{R^{2}}}    ~,\eea
in the left-hand side (LHS) of the B-ball equation, we obtain
\bea{a8}  \phi^{''} + \frac{2}{r} \phi^{'} = \left(- \frac{6}{R^{2}} + 
\frac{4 r^{2}}{R^{4}}
\right) \phi 
~.\eea
Inserting in the RHS gives 
\bea{a9}  - \omega_{o}^{2} \phi  - 
 \left( \frac{2 K m^{2}}{R^{2}} \right) r^{2} \phi     \equiv  (A + B r^{2}) \phi   ~,\eea
where we have set $M = \phi(0)$.
Thus we see that the same form is obtained on the LHS and RHS. The exponential factor in the Gaussian will be 
correct up to a factor of the order of 1 if the values of $A$ and $B$ can be consistently approximated by a single value of 
$R$. ($r$ as a function of $\phi$ will be given correctly up to a factor 
of the order of the square root of the factor between the LHS and RHS). 
This requires that $\omega_{o}^{2} \approx 3 |K|m^{2}$ and $R^{2} \approx 
2 (|K|m^{2})^{-1}$ in the numerical solution. 

	 Let us note some other properties of the Gaussian B-ball Ansatz. 
The total charge of the B-ball is given by 
\bea{a10} B = \int dr \; 4\pi r^{2} \omega \phi_{o}^{2} e^{- \frac{2 r^{2}}{R^{2}} } 
 = \left(\frac{\pi}{2}\right)^{3/2} \omega \phi_{o}^{2} R^{3}    ~.\eea
The total energy of the B-ball is given by
\bea{a11} E \approx \frac{3}{2} \left(\frac{\pi}{2}\right)^{3/2} \phi_{o}^{2} R + 
\left(\frac{\pi}{2}\right)^{3/2} m^{2} \phi_{o}^{2} R^{3}       ~,\eea
where the second term is the combined contribution from the potential 
energy and the charge term and we have used $\omega^{2} \approx m^{2}$.
 Since $R$ is large compared with $m^{-1}$ for small $|K|$, 
the potential plus charge term dominates the energy. The radius within which 90$\%$ of the
energy is found, $R_{c}$, is then given by $R_{c} = 1.25 R$.
The energy per unit charge is given by
\bea{a11a} \frac{E}{B} = \frac{m^{2}}{\omega} \approx 
\left(1 + \frac{3 |K|}{2}\right) m    ~,\eea
where we have used the Gaussian result $\omega_{o}^{2} = 3 |K|m^{2}$. Although the 
energy per unit charge is larger than $m$, the mass of the 
scalar at small values of $\phi$ will be of the form $m (1 + \alpha |K|)$ (with $\alpha \gae 1$)
once the logarithmic correction to the potential is included, 
so that the binding energy per unit charge will be positive and of the order of $|K| m$. 

	Comparing the Gaussian solution with the numerical solution, using the 
numerical value of $R_{c}$ to normalize the Gaussian solution, we obtain  
$R = k_{R} (|K|^{1/2}m)^{-1}$ with $k_{R} \approx 1.2$, in good agreement       
with the value expected from the Gaussian Ansatz. The value of $\omega_{o}$ 
from the numerical solution is about a factor of two larger than that expected from the 
Gaussian Ansatz. Thus we expect that $r$ as a function of $\phi$ will be correctly given by the
Gaussian Ansatz  up to a factor of about 2.  

		  Physically, a factor of 2 variation in $\omega_{o}$ correspond to a variation in 
$|K|^{1/2}$ or $m$ by a factor of 2 in the
 real solution, which will not greatly alter the physical properties of
 the corresponding B-ball. Thus we expect the Gaussian to be a physically reasonable 
approximation to the B-ball solution.

\end{document}